\documentstyle[aps,epsfig,floats,eqsecnum,prd]{revtex}

\def\laq{\raise 0.4ex\hbox{$<$}\kern -0.8em\lower 0.62ex\hbox{$\sim$}}
\def\gaq{\raise 0.4ex\hbox{$>$}\kern -0.7em\lower 0.62ex\hbox{$\sim$}}

\newcommand{\beq}{\begin{equation}} 
\newcommand{\eeq}{\end{equation}}
\newcommand{\bea}{\begin{eqnarray}} 
\newcommand{\eea}{\end{eqnarray}}
\newcommand{\ba}{\begin{array}} 
\newcommand{\ea}{\end{array}}

\begin{document}
\draft
\title{\large\bf Quantum noise in second generation, signal-recycled  
laser interferometric gravitational-wave detectors}
\author{Alessandra Buonanno and Yanbei Chen}
\address{Theoretical Astrophysics and Relativity Group,\\
California Institute of Technology,
Pasadena, CA 91125, USA }
\vskip 0.2truecm
\maketitle
\begin{abstract}
It has long been thought that the sensitivity of 
laser interferometric gravitational-wave detectors is limited by 
the free-mass standard quantum limit, unless radical redesigns of 
the interferometers or modifications of their input/output 
optics are introduced. Within a fully quantum-mechanical 
approach we show that in a second-generation interferometer  
composed of arm cavities and a signal recycling cavity, e.g., the 
LIGO-II configuration, (i) quantum shot noise and quantum 
radiation-pressure-fluctuation noise are dynamically correlated, 
(ii) the noise curve exhibits two resonant dips, 
(iii) the Standard Quantum Limit can be beaten by a factor of 2, 
over a frequency range $\Delta f/f \! \sim \! 1$,  but at the price 
of increasing noise at lower frequencies. 
\end{abstract}
\vskip 0.2truecm
\centerline{\small  PACS No.: 04.80.Nn, 95.55.Ym, 42.50.Dv, 03.65.Bz.
\hspace{0.5 cm} GRP/00/553}
\section{Introduction}
\label{sec1}

Several laser interferometric gravitational-wave (GW) detectors~\cite{Inter} 
(interferometers for short), sensitive to the high-frequency band 
$10-10^3 \, {\rm Hz}$, will become operative within about one year. 
In the first generation of these interferometers the Laser Interferometer 
Gravitational Wave Observatory (LIGO), TAMA and Virgo  configurations
\footnote{~GEO's optical configuration 
differs from that of LIGO/TAMA/Virgo --- it does not have Fabry-Perot
cavities in its two Michelson arms, and the analysis made in this paper does not
directly apply to it. However, we note that GEO, already in its first implementation,
does use the `signal recycling' optical configuration with which this paper
deals. }
are characterized by kilometer-scale arm 
cavities  with four mirror-endowed test masses, suspended from seismic-isolation stacks. 
Laser interferometry is used to monitor the relative change in the
positions of the mirrors 
induced by the gravitational waves.
The Heisenberg uncertainty principle, applied to 
the test masses of GW interferometers states that, if the relative positions 
are measured with high precision, then the test-mass momenta 
will be perturbed.  As time passes, the momentum perturbations will produce 
position uncertainties, which might mask the tiny displacements produced 
by gravitational waves. If the momentum perturbations and measurement errors
are not correlated, a detailed analysis of the above process gives 
rise to the standard quantum limit (SQL) for interferometers: a limiting (single-sided)
noise spectral density $S_h^{\rm SQL} = 8 \hbar/(m\Omega^2L^2)$ for the 
dimensionless gravitational-wave signal $h(t) = \Delta L/L$ \cite{KT80}.
Here $m$ is the mass of each identical test mass, $L$ is the length of 
the interferometer's arms, $\Delta L$ is the time evolving difference in the 
arm lengths, $\Omega$ is the GW angular frequency, and $\hbar$ 
is Planck's constant.  

The concept of SQL's for high-precision measurements was first formulated 
by Braginsky~\cite{B68-70s}. He also demonstrated that it 
is possible to circumvent SQL's by changing the designs of the 
instruments, so they measure quantities which are not 
affected by the uncertainty principle by virtue of commuting 
with themselves at different times \cite{B68-70s}, \cite{BK92} -- as for example in 
speed-meter interferometers \cite{SM}, which measure test-mass momenta  
instead of positions.
Interferometers that circumvent the SQL are called quantum-nondemolition 
(QND) interferometers.  
Since the early 1970s, it has been thought that 
to beat the SQL for GW interferometers the redesign must be major. 
Examples are (i) speed-meter designs \cite{SM} with their radically 
modified optical topology, (ii) the proposal to inject squeezed vacuum 
into an interferometer's dark port \cite{SFD}, and (iii) the proposal 
to introduce two kilometer-scale filter cavities into the interferometer's 
output port~\cite{KLMTV00} so as to implement frequency-dependent 
homodyne detection \cite{HFD}. Both (ii) and (iii) intend to take
advantage of the non-classical correlations of the optical fields. These
radical redesigns require high laser power circulating in the 
arm cavities  ($\gaq \, 1$ MW)  and/or are 
strongly susceptible to optical losses which tend to destroy quantum correlations.      
In order to tackle these two important issues, Braginsky, Khalili and 
colleagues have recently proposed the GW ``optical bar'' scheme \cite{OB},
where the test mass is effectively an oscillator, whose restoring 
force is provided by in-cavity optical fields.
For ``optical bar'' detectors the free-mass
SQL is no more relevant and one can beat the SQL using classical 
techniques of position monitoring.
 Moreover, this scheme has two major advantages: It requires 
much lower laser power circulating in 
the cavities \cite{OB}, and is less 
susceptible to optical losses.

Research has also been carried out using successive independent
monitors of free-mass positions. Yuen, Caves and Ozawa discussed 
and disputed about the applicability and the beating of the SQL 
within such models \cite{debate}. Specifically, Yuen and Ozawa conceived ways to
beat the SQL by taking advantage of the so-called contractive states
\cite{debate}. However, the class of interaction Hamiltonians given by Ozawa are 
not likely to be applicable to GW interferometers (for further details 
see Ref. \cite{BC300}).

Recently, we showed \cite{BC100} that it is possible 
to circumvent the SQL for LIGO-II-type signal-recycling (SR)
interferometers  \cite{SR,M95}.
With their currently planned design, LIGO-II interferometers can beat the SQL by modest 
amounts, roughly a factor two over a bandwidth 
$\Delta f \!\sim\! f$.
\footnote{~{\it If} all sources of thermal noise can also be
pushed below the SQL.
The thermal noise is a tough problem and  
for current LIGO-II designs with $30$ kg sapphire mirrors,
estimates place its dominant, thermoelastic component slightly above
 the SQL \cite{BGV00}. } 
It is quite interesting to notice that the beating of the SQL in 
SR interferometers has a similar origin as in  
``optical bar'' GW detectors mentioned above \cite{OB}.

{}Braginsky and colleagues \cite{BGKMTV00}, building on earlier 
work of Braginsky and Khalili \cite{BK92}, have shown that for 
LIGO-type GW interferometers, the test-mass initial quantum state only 
affects frequencies $ \,\laq\, 1$\,Hz, the dependence on 
the initial quantum state can be removed filtering the 
output data at low frequency. 
Therefore, the SQL in GW interferometers is enforced only 
by the light's quantum noise, {\it not directly} by the test mass.
As we discussed in \cite{BC100}, and we shall explicitly show below,
we can decompose the optical noise of a SR interferometer 
into shot noise and radiation-pressure noise, using the fact that
they transform differently under rescaling 
of the mirror mass $m$ and the light power $I_o$.  
As long as there are no correlations between 
the light's shot noise and its radiation-pressure-fluctuation noise, 
the light firmly enforces the SQL. This is the case 
for conventional interferometers, i.e.\ for interferometers that
have no SR mirror at the output dark port and a simple homodyne
detection is performed (the type of
interferometer used in LIGO-I/TAMA/Virgo).  
However, the SR mirror \cite{SR,M95} 
(which is being planned for LIGO-II\,\footnote{~
The LIGO-II configuration will also use a power-recycling cavity 
to increase the light power at the beamsplitter.  
The presence of this extra cavity will not affect the 
quantum noise in the dark-port output. For this reason 
we do not take it into account.} 
as a tool to reshape the noise curve, and thereby improve the 
sensitivity to specific GW sources \cite{Sour}
) 
produces \emph{dynamical} shot-noise / back-action-noise 
correlations, and {\it these correlations break the light's 
ability to enforce the SQL}. 
These dynamical correlations come naturally from the 
nontrivial coupling between the test mass and the signal-recycled optical fields,
which makes the dynamical properties of the entire optical-mechanical 
system rather different from the naive picture of a free mass buffeted 
by Poissonian radiation pressure. 
As a result, the SQL for a free test mass has no relevance 
for a SR interferometer. Its only remaining role is as a 
reminder of the regime where back-action
noise is comparable to the shot noise.
The remainder of this paper is devoted to explaining these claims 
in great detail. To facilitate the reading we 
have put our discussion  of the dynamical system 
formed by the optical fields 
and the mirrors into a separate, companion paper~\cite{BC300}.

The outline of this paper is as follows. In Sec.~\ref{sec2} we derive  
the input--output relations for the whole optical system composed  
of arm cavities and a SR cavity, pointing out the existence 
of dynamical instabilities, and briefly commenting on the possibility 
and consequences of introducing a control system to suppress them.
In Sec.~\ref{sec3} we evaluate the spectral density of the quantum noise. 
More specifically, in Sec.~\ref{subsec3.1} we discuss the general case, showing 
that LIGO-II  can beat the SQL when 
dynamical correlations between shot noise and radiation-pressure 
noise are produced by the SR mirror.
In Sec.~\ref{subsec3.2}, making links to previous investigations, 
we decompose our expression for the optical noise into shot noise and 
radiation-pressure noise and express the dynamical correlations between the two noises
in terms of physical parameters characterizing the SR interferometer;
in Sec.~\ref{subsec3.3} we specialize to two cases, 
the extreme signal-recycling (ESR) and extreme resonant-sideband-extraction 
(ERSE) configurations, where dynamical correlations are absent 
and a semiclassical approach can be applied \cite{SR,M95}. 
In Sec.~\ref{sec4} we investigate the structure of resonances
of the optical-mechanical system and discuss their link to the minima present
in the noise curves.
Finally, Sec.~\ref{sec5} deals with the effects of optical losses, while 
Sec.~\ref{sec6} summarizes our main conclusions. The Appendix  discusses 
the validity of the two-photon formalism in our context. 

\section{Signal-recycling interferometer: input--output relations} 
\label{sec2}

\begin{figure}
\begin{center}
\vspace{-0.5cm} 
\epsfig{file=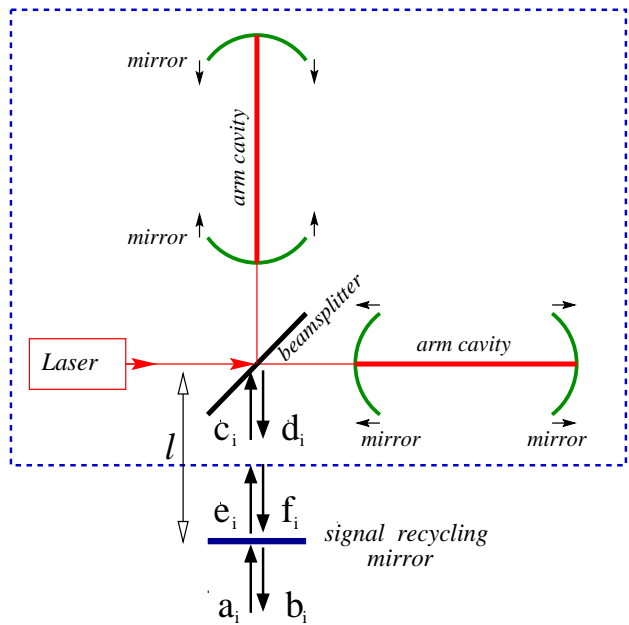,width=0.5\textwidth,height = 0.5\textwidth,angle=-90}
\vskip 0.2truecm
\caption{ 
Schematic view of a {\rm LIGO-II} signal-recycling interferometer. 
The interior of the dashed box refers to the conventional 
interferometer; $c_i$ and $d_i$ are the input and output fields
at the beam splitter's dark port; $a_i$ and $b_i$ are the 
vacuum input and signal output of the whole optical 
system. The laser light enters the bright port of the beam splitter. 
The arrows close to arm cavities' extremities 
indicate gravitational-wave-induced mirror displacements.}
\label{Fig1}
\vspace{-0.6cm} 
\end{center}
\end{figure}

In Fig.~\ref{Fig1} we sketch the SR configuration of LIGO-II interferometers. 
The optical topology inside the dashed box is that of conventional interferometers 
such as LIGO-I/TAMA/Virgo, which are Michelson interferometers 
with Fabry-Perot (FP) arm cavities. The principal noise input and 
the signal and noise output for the conventional topology are 
$c_i$ and $d_i$  in Fig.~\ref{Fig1}.
In a recent paper, KLMTV \cite{KLMTV00} have derived the 
input--output $(c_i - d_i)$ relations for a conventional interferometer at the 
output dark port, immediately after the beam splitter, within a full 
quantum mechanical approach. 
In this section we shall derive the input--output $(a_i - b_i)$ relations for 
the whole optical system at the output port, i.e.\ immediately 
after the SR mirror, and shall evaluate the corresponding noise spectral
density. 

As we shall see, a naive application of the Fourier-based formalism 
developed in \cite{KLMTV00} gives ill-defined input--output relations, due to the presence 
of optical-mechanical instabilities. 
These instabilities have an origin similar to the dynamical instability 
of a detuned FP cavity induced by  the radiation-pressure force 
acting on the mirrors, which has long been investigated 
in the literature \cite{All,HR00,PDHV00}. 
To suppress the growing modes and make the KLMTV's formalism valid 
for SR interferometers, an appropriate control system should be introduced. 
The analysis of the resulting interferometer plus controller 
requires a detailed description of the dynamics of the whole 
system and for this we have found Braginsky and Khalili's theory of linear quantum 
measurement \cite{BK92} very powerful and intuitive.
We analyzed the details of the dynamics in an accompanying paper
\cite{BC300}, showing in particular that the results derived in 
this section by Fourier techniques, notably the  noise spectral 
density curves, are correct and rigorously justified.

\subsection{Naive extension of KLMTV's results to SR interferometers}
\label{subsec2.1}
As in Ref.~\cite{KLMTV00} we shall describe the interferometer's light by 
the electric field evaluated on the optic axis 
(center of light beam) and at specific, fixed locations along the 
optic axis. Correspondingly, the electric fields that we write 
down will be functions of time only: all dependence on spatial 
position will be suppressed from our formulae.
\begin{table}
\squeezetable
\begin{tabular}{cccc}
{\sl Quantity} & {\sl Symbol \& value for {\rm LIGO-II}} 
& {\sl Quantity} & {\sl Symbol \& value for {\rm LIGO-II}} \\\hline \hline
Light power at beam splitter & $I_o$ &
Light power to reach SQL & $I_{\rm SQL} = 1.0 \times 10^4$ {\rm W} \\
SQL for GW detection & $h^2_{\rm SQL} \equiv S_h^{\rm SQL} = 4 \times 10^{-48}/{\rm Hz}$ &
Arm-cavity half bandwidth & $\gamma = Tc/4L= 2\pi \times 100\,{\rm sec}^{-1}$ \\
Laser angular frequency & $\omega_o = 1.8 \times 10^{15}\,{\rm sec}^{-1}$ &
GW angular frequency & $\Omega$ \\
End-mirror mass & $m= 30$ {\rm kg} &
Arm-cavity length & $L = 4$ {\rm km} \\ 
SR cavity length & $l \approx 10$ {\rm m} &
Internal arm-cavity mirror transmissivity & $T = 0.033$ (power)\\
SR mirror transmissivity & $\tau$ (amplitude)&
SR cavity detuning  & $\phi$ \\
Arm-cavity power loss & $\epsilon = 0.01$ & 
SR power loss & $\lambda_{\rm SR} = 0.02$ \\ 
Photodetector loss & $\lambda_{\rm PD} = 0.1$&  
\end{tabular}
\caption{\sl Summary of {\rm LIGO-II} parameters \protect\cite{GSSW99}.}
\label{Tab1}
\end{table}

The input field at the bright port 
of the beam splitter, which is assumed 
to be infinitesimally thin, is a carrier field, described by a coherent 
state with power $I_o$ and angular frequency $\omega_o$. 
We assume \cite{KLMTV00} that the arm-cavity end mirrors 
oscillate around an equilibrium position that is on resonance 
with the carrier light. This means that there is no zeroth-order arm-cavity detuning  
(see the paper of Pai et al.\ \cite{PDHV00} for a critical 
discussion of this assumption).
Our most used interferometer parameters are given in Table~\ref{Tab1} 
together with the values anticipated for LIGO-II.

We denote by $f_{\rm GW}=\Omega/2 \pi$ the GW frequency, which lies in 
the range $10-1000$ Hz. Then the interaction of a gravitational wave with 
the optical system produces side-band frequencies $\omega_o \pm \Omega$ 
in the electromagnetic field at the output dark port. For this reason, 
similarly to KLMTV~\cite{KLMTV00}, we find it convenient to describe the quantum 
optics inside the interferometer using the two-photon formalism 
developed by Caves and Schumaker \cite{CS185,CS285}. In this formalism,  
instead of using the usual annihilation and creation operators for photons 
at frequency $\omega$, we expand the field operators in terms of quadrature operators 
which can simultaneously annihilate a photon at 
frequency $\omega = \omega_o + \Omega$ while creating a photon at 
frequency $\omega = \omega_o - \Omega$ 
(or vice versa). 

More specifically, the quantized electromagnetic field in the Heisenberg picture 
evaluated at some fixed point on the optic axis, and restricted to the 
component propagating in one of the two directions along the axis is:
\beq
\label{2.1}
\widehat{E}(t) = \sqrt{\frac{2 \pi \hbar}{{\cal A}\,c}}\,
\int_0^{+\infty} \sqrt{\omega}\,\left [ \widehat{a}_\omega\,e^{-i\omega t}
+ \widehat{a}^\dagger_\omega\,e^{+i\omega t} \right ]\,\frac{d \omega}{2 \pi}\,.
\eeq
Here ${\cal A}$ is the effective cross sectional area of the laser beam and 
$c$ is the speed of light. The annihilation and creation operators 
$\widehat{a}_\omega$, 
$\widehat{a}_\omega^\dagger$ in Eq.~(\ref{2.1}), which in the Heisenberg picture are 
fixed in time, satisfy the usual commutation relations
\beq
[\widehat{a}_\omega, \widehat{a}_{\omega'}] =0\,, \quad \quad 
[\widehat{a}^\dagger_\omega, \widehat{a}^\dagger_{\omega'}] =0\,, \quad \quad 
[\widehat{a}_\omega, \widehat{a}_{\omega'}^\dagger] = 2 \pi\,\delta(\omega - 
\omega^\prime)\,.
\label{2.2}
\eeq

Henceforth, to ease the notation we shall omit the hats on quantum 
operators. Defining the new operators (see Sec.~IV of Ref.~\cite{CS185}
\footnote{~
Our notations are not exactly the same as those of Caves and Schumaker 
\protect\cite{CS185,CS285}, the correspondence is the following 
(ours $\rightarrow$ Caves-Schumaker):
$ \omega_0 \rightarrow \Omega, \,\,
\Omega \rightarrow \epsilon, \,\,
a_{\omega_0 \pm \Omega} \rightarrow a_{\pm},\,\,
a_{\pm} \rightarrow \lambda_{\pm}a_{\pm}, \,\,
a_{1,2} \rightarrow \alpha_{1,2}$.
We refer to Sec.~IV B of \protect\cite{CS185} for further details.})
\beq
a_+ \equiv a_{\omega_o + \Omega}\,\sqrt{\frac{\omega_o + \Omega}{\omega_o}} \,,
\quad \quad 
a_- \equiv a_{\omega_o - \Omega}\,\sqrt{\frac{\omega_o - \Omega}{\omega_o}}\,,
\label{2.3}
\eeq
and using the commutation relations (\ref{2.2}), we find 
\bea
&& [a_+, a_{+^\prime}^\dagger] = 2 \pi\,\delta(\Omega - \Omega^\prime)\,
\left ( 1 + \frac{\Omega}{\omega_o} \right )\,, \quad \quad 
[a_-, a_{-^\prime}^\dagger] = 2 \pi\,\delta(\Omega - \Omega^\prime)\,
\left ( 1 - \frac{\Omega}{\omega_o} \right )\,, \\
\label{2.4}
&& [a_+, a_{+^\prime}] = 0 =  [a_-, a_{-^\prime}]\,, \quad \quad 
[a_+^\dagger, a_{+^\prime}^\dagger] = 0 =  [a_-^\dagger, a_{-^\prime}^\dagger]\,,
\quad \quad  [a_+, a_{-^\prime}] = 0 = [a_+, a_{-^\prime}^\dagger]\,,
\label{2.5}
\eea
where $a_{\pm^\prime}$ stands for $a_{\pm}(\Omega^\prime)$.
Because the carrier frequency is $\omega_o \simeq 10^{15}\,\rm s^{-1}$ and we are interested in 
frequencies $\Omega/2 \pi$ in the range $10\,--\,10^3$ Hz, we shall disregard in Eq.~(\ref{2.4}) 
the term proportional to $\Omega/\omega_o$. (In the Appendix  we shall give 
a more complete justification of this by evaluating the effect the term 
proportional to  $\Omega/\omega_o$ would have on the final noise spectral density.)
We can then rewrite the electric field, Eq.~(\ref{2.1}), as 
\beq
\label{2.6}
E(t) = \sqrt{\frac{2 \pi \hbar\,\omega_o}{{\cal A}\,c}}\,
e^{-i\omega_o\,t}\,\int_0^{+\infty} 
(a_+(\Omega)\,e^{-i\Omega t} + a_-(\Omega)\,e^{i\Omega t})\,
\frac{d \Omega}{2 \pi}  + {\rm h.c.}\,,
\eeq
where ``h.c.'' means Hermitian conjugate. 
Following the Caves-Schumaker two-photon formalism~\cite{CS185,CS285}, we introduce 
the amplitudes of the two-photon modes as  
\beq
\label{2.7}
a_1 = \frac{a_+ + a_-^\dagger}{\sqrt{2}}\,, \quad \quad 
a_2 = \frac{a_+ - a_-^\dagger}{\sqrt{2}i}\,;
\eeq
$a_1$ and $a_2$ are called quadrature fields and they satisfy the 
commutation relations
\bea
\label{2.8}
&& [a_1, a_{2^\prime}^{\dagger}] = - [a_2, a_{1^\prime}^{\dagger}]=
2\pi i \delta(\Omega-\Omega^\prime)\,, \nonumber \\
&& [a_1, a_{1^\prime}^{\dagger}] = 0= [a_1, a_{1^\prime}]\,,
\quad \quad 
[a_2, a_{2^\prime}^{\dagger}] = 0= [a_2, a_{2^\prime}]\,.
\label{com}
\eea
Expressing the electric field (\ref{2.6}) in terms of the 
quadratures we finally get
\beq 
\label{2.9}
E(a_i;t) = \cos (\omega_o\,t)\,E_1(a_1;t) + \sin (\omega_o\,t)\,E_2(a_2;t)\,,
\eeq
with 
\beq
\label{2.10}
E_j(a_j;t) = \sqrt{\frac{4 \pi \hbar\,\omega_o}{{\cal A}\,c}}\,
\int_0^{+\infty} (a_j\,e^{-i\Omega t} + a_j^{\dagger}\,e^{i\Omega t})\,
\frac{d \Omega}{2 \pi} \quad \quad j = 1,2\,.
\eeq
Note (as is discussed at length by BGKMTV \cite{BGKMTV00} and 
was previewed by KLMTV (footnote 1 of Ref.~\cite{KLMTV00})),
that, $E_1(t)$ and $E_2(t)$ commute with themselves 
at any two times $t$ and $t'$, i.e.\ $[E_j(t),E_j(t^\prime)]=0$, 
while $[E_1(t),E_2(t^\prime)] \! \sim \! i\delta(t-t')$.
Hence, the quadrature fields $E_j(t)$ with $j=1,2$ are quantum-nondemolition 
quantities which can be measured with indefinite accuracy over time, i.e.\ 
measurements made at different times can be stored as independent bits of data 
in a classical storage medium without being affected by mutually-induced noise, while 
it is not possible to do this for $E_1(t)$ and $E_2(t)$ simultaneously. 
As BGKMTV \cite{BGKMTV00} emphasized (following earlier work by Braginsky and 
Khalili \cite{BK92}), this means that we can regard $E_1(t)$ and $E_2(t)$ 
separately as classical variables -- though in each other's presence they 
behave nonclassically.

For GW interferometers the full input electric field at the 
dark port is $E(c_i; t)$ where $c_1$ and $c_2$ are the two input quadratures,    
while the output field at the dark port is 
$E(d_i;t)$, with $d_1$ and $d_2$ the two output quadratures (see Fig.~\ref{Fig1}). 
Assuming that the classical laser-light input field at the beam splitter's bright 
port is contained only in the first quadrature,\footnote{~For the KLMTV 
optical configuration and for ours, only a negligible fraction 
of the quantum noise entering the bright port emerges from the 
dark port.} and evaluating the back-action force acting on the 
arm-cavity mirrors disregarding the motion of the mirrors during the 
light round-trip time (quasi-static approximation),
\footnote{~The description of a SR interferometer beyond the quasi-static approximation 
\protect\cite{PDHV00,HR00} introduces nontrivial corrections to the back-action force,  
proportional to the power transmissivity $T$ of the input arm-cavity mirrors. 
Since $T \simeq 0.033$ (see Table \ref{Tab1}) we expect a small 
modification of our results, but an explicit calculation is 
strongly required to quantify these effects.} 
KLMTV~\cite{KLMTV00} derived the following input--output relations 
at side-band (GW) angular frequency $\Omega$ 
\beq
\label{2.11}
d_1 = c_1\, e^{2 i \beta}\,, \quad \quad d_2=(c_2-{\cal K} c_1)\, 
e^{2 i \beta}+\sqrt{2{\cal K}}\, \frac{h}{h_{\rm SQL}}\,e^{i \beta}\,,
\eeq
where $2\beta=2\arctan{{\Omega}/{\gamma}}$ is the net phase 
gained by the sideband frequency $\Omega$ while in the arm-cavity, 
$\gamma = Tc/4L$ is the half bandwidth of the arm-cavity 
($T$ is the power transmissivity of the arm-cavity input mirrors and 
$L$ is the length of the arm cavity), $h$ is the Fourier 
transform of the gravitational-wave field, and $h_{\rm SQL} $ 
is the SQL for GW detection, explicitly given by
\beq
h_{\rm SQL}(\Omega)\equiv \sqrt{S_h^{\rm SQL}}=\sqrt{\frac{8 \hbar}{m \Omega^2 L^2}}\,,
\label{2.12}
\eeq
where $m$ is the mass of each arm-cavity mirror. The quantity ${\cal K}$ in Eq.~(\ref{2.11}) 
is the effective coupling constant, which relates 
the motion of the test mass to the output signal,
\beq
{\cal K}=\frac{2 (I_o/I_{\rm SQL}) \gamma^4}{\Omega^2(\gamma^2+\Omega^2)}\,. 
\label{2.13}
\eeq 
Finally, $I_o$ is the input light power, and $I_{\rm SQL}$ is the light power 
needed by a conventional interferometer to reach the SQL at a side band 
frequency $\Omega=\gamma$, that is
\beq
I_{\rm SQL} = \frac{m\,L^2\,\gamma^4}{4 \omega_o}\,.
\label{2.14}
\eeq
(See in Table~\ref{Tab1} the values of the interferometer 
parameters tentatively planned for LIGO-II \cite{GSSW99}.)
We shall now derive the {\it new} input--output $(a_i - b_i)$ 
relations including the SR cavity.
We indicate by $l$ the length of the SR cavity
and we introduce two dimensionless 
variables: $\phi\equiv [\omega_o l/c]_{{\rm mod}\, 2\pi}$, \footnote{~Note that
$\omega_o l/c = 2\pi m + \phi$, with $m$ a large integer. Indeed,  
typically $\omega_o \simeq 10^{15}\,{\rm s^{-1}}$, $l \simeq 10 \,{\rm m}$, hence  
$\omega_o l/c \gg 1$.} the phase gained by the carrier frequency $\omega_o$ 
while traveling one way in the SR cavity, and 
$\Phi\equiv[\Omega l/c]_{{\rm mod}\, 2\pi}$ the additional phase 
gained by the sideband with GW frequency $\Omega$ (see Fig.~\ref{Fig1}). 
Note that we are assuming that the distances from the 
beam splitter to the two arm-cavity input mirrors are identical, equal to
an integer multiple of the carrier light's wavelength, and are 
negligible compared to $l$. 

Propagating the output electric field 
$E(d_i;t)$ up to the SR mirror, 
and introducing the operators $e_i$ and $f_i$ which 
describe the fields that are immediately inside the SR 
mirror (see Fig.~1), we get the condition 
\beq
\label{2.15}
E(f_i; t) 
= E\left (d_i;t - \frac{l}{c} \right )\,,
\eeq
which, together with Eq.~(\ref{2.9}), provides the following equations
\beq
\label{2.16}
f_1 = (d_1\,\cos{\phi}-d_2\,\sin{\phi})\,e^{i\Phi}\,, 
\quad \quad f_2 = (d_1\,\sin{\phi}+d_2\,\cos{\phi})\,e^{i\Phi}\,.
\eeq
Proceeding in an analogous way for the input electric field $E(c_i;t)$,      
we derive
\beq
\label{2.17}
e_1 = (c_1\,\cos{\phi}+c_2\,\sin{\phi})\,e^{-i\Phi}\,, \quad \quad 
e_2 =(-c_1\,\sin{\phi}+c_2\,\cos{\phi})e^{-i\Phi}\,.
\eeq
Note that each of Eqs.~(\ref{2.16}), (\ref{2.17}) correspond to  
a rotation of the quadratures $d_1$, $d_2$ 
(or $c_1$, $c_2$) plus the addition of an overall phase.
Finally, denoting by $a_i$ and $b_i$ the input and output fields 
of the whole system at the output port (see Fig.~\ref{Fig1}) 
we conclude that the following relations should be satisfied at the SR mirror: 
\bea
\label{2.18}
&& e_1 = \tau\,a_1+\rho\, f_1 \,, \quad \quad 
e_2 = \tau\, a_2+\rho\, f_2 \,,\\
\label{2.19}
&& b_1= \tau\,f_1-\rho\,a_1\,, \quad \quad \,\,
b_2 = \tau\, f_2-\rho\, a_2\,,
\eea
where $\pm\rho$ and $\tau$ are the amplitude reflectivity and transmissivity 
of the SR mirror, respectively. We use the convention that 
$\rho$ and $\tau$ are real and positive, with the reflection coefficient  
being $+\rho$ for light coming from inside the cavity and $-\rho$ 
for light coming from outside. In this section we limit ourselves 
to a lossless SR mirror; therefore the 
following relation holds: $\tau^2 + \rho^2 =1$.

Before giving the solution of the above equations, let us notice that 
the equations we derived so far for the quantum EM fields 
in the Heisenberg picture are exactly the same as those of classical EM fields. 
To deduced them it is sufficient 
to replace the quadrature operators  by the Fourier
components of the classical EM fields.
The input-output relation we shall give below is also the same as 
in the classical case. In the latter we  
should assume that a fluctuating field enters the input port of the entire 
interferometer. More specifically, assuming a vacuum state in the input port, we can
model the two input quadrature fields as two independent white noises. Then 
using the classical equations, we can derive the
output fields which have the correct noise spectral densities.

Solving the system of Eqs.~(\ref{2.11}), (\ref{2.16})--(\ref{2.19}) 
gives the final input--output relation:
\beq
\left (\matrix{b_1 \cr b_2}\right)=
\frac{1}{M}\left[e^{2i(\beta+\Phi)}\left(\matrix{ C_{11}& C_{12}\cr C_{21} & C_{22}}\right)
\left(\matrix{a_1 \cr a_2 }\right)+
\sqrt{2 {\cal K}}\tau
e^{i(\beta+\Phi)}
\left(\matrix{D_1 \cr D_2} \right)\frac{h}{h_{\rm SQL}}\right]\,,
\label{2.20}
\eeq
where, to ease the notation, we have defined: 
\beq
M= 1 + \rho^2\, e^{4i(\beta+\Phi)}-
2\rho\,e^{2i(\beta+\Phi)}\left (\cos{2\phi}+\frac{{\cal K}}{2}\,
\sin{2\phi} \right )\,,
\label{2.21}
\eeq
\bea
\label{2.22}
&& C_{11}=C_{22}=(1+\rho^2)\,
\left (\cos{2\phi}+\frac{{\cal K}}{2}\,\sin{2\phi} \right ) -2\rho\,\cos{(2(\beta+\Phi))}\,,\\
&& C_{12}=-\tau^2\,(\sin{2\phi}+{\cal K}\,\sin^2{\phi})\,, \quad \quad 
C_{21}=\tau^2\,(\sin{2\phi}-{\cal K}\,\cos^2{\phi})\,,
\label{2.23}
\eea
\beq
D_1= - (1+\rho\, e^{2 i(\beta+\Phi)})\,\sin{\phi}\,, \quad \quad 
D_2= - (-1+\rho\, e^{2 i(\beta+\Phi)})\,\cos{\phi}\,.
\label{2.24}
\eeq
A straightforward calculation using $C_{ij} \in \Re$ and $C_{11}\,C_{22} 
- C_{12}\,C_{21} = |M|^2$, confirms that the quadratures $b_i, b_i^\dagger$ 
satisfy the commutation relations (\ref{2.8}), as they should since like 
$a_i$ and $a_i^\dagger$ they represent free fields. 
Let us also observe that both the quadratures $b_1$ and $b_2$ in Eq.~(\ref{2.20}) 
contain the gravitational-wave signal $h$ and that it is not possible 
to put the signal into just one of the quadratures through a transformation that 
preserves the commutation relations of $b_1$ and $b_2$. Indeed, 
the most general transformation that preserves the commutation relations is of the form 
\beq
\left(\matrix{\widetilde{b}_1 \cr \widetilde{b}_2}\right)=
e^{i\,\alpha}\,\left(\matrix{ L_{11} & L_{12}\cr L_{21} & L_{22}}\right)
\left(\matrix{ b_1 \cr b_2}\right)\,, \quad \quad 
L_{i j} \in \Re\,, \quad \det{L_{i j}} =1\,,
\label{2.25}
\eeq 
where $\alpha$ is an arbitrary phase. Because the $D_i$ are complex 
[see Eq.~(\ref{2.24})], it is impossible to null 
the $h$ contribution either in $\widetilde{b}_1$ or $\widetilde{b}_2$.

Henceforth, we limit our analysis to $\Phi=0$, which corresponds to 
a SR cavity much shorter than the arm-cavities, e.g., $l \simeq 10\,{\rm m}$.
We assume for simplicity that there is no radio-frequency (MHz)  
modulation/demodulation of the carrier and the signal \cite{GSSW99};
instead, some frequency-independent quadrature 
\bea
b_\zeta&&=b_1\,\sin\zeta+b_2\,\cos\zeta \nonumber\\
&&=\frac{1}{M}\,\left [
e^{2 i \beta}
(C_{11}\,\sin\zeta+C_{21}\,\cos\zeta)\,a_1+
e^{2 i \beta}
(C_{12}\,\sin\zeta+C_{22}\,\cos\zeta)\,a_2 \right . \nonumber \\ 
&&\left .  + 
\sqrt{2{\cal K}}\,\tau\,e^{i \beta}\,(D_1\,\sin\zeta+D_2\,\cos\zeta)\,
\frac{h}{h_{\rm SQL}} \right ]\,,
\label{2.26}
\eea
is measured via homodyne detection \cite{HFD}. 
\footnote{~It is still unclear what
detection scheme (direct homodyne detection or RF modulation/demodulation ) 
will be used in LIGO-II. The decision will require a 
quantum-mechanical analysis of the
additional noise introduced by the modulation/demodulation process, which 
will be given in a future paper \cite{BC400}.}
Before going on to evaluate the noise spectral density in the measured 
quadrature $b_{\zeta}$, let us first comment on the results 
obtained in this section.

\subsection{Discussion of the naive result}
\label{subsec2.2}
There is a major delicacy in the input--output relation given by Eq.~(\ref{2.18}). 
By naively transforming it from the frequency domain 
back into the time domain, we deduce that the output 
quadratures depend on the gravitational-wave field and the input optical 
fields both in the past \emph{and in the future}. 
Mathematically this is due to the fact that the coefficient $1/M$, in front of $h$ 
and $a_i$ ($i=1,2$) in Eq.~(\ref{2.20}), contains poles both in the lower 
\emph{and in the upper} complex plane. This situation is a very common one 
in physics and engineering (it occurs for example 
in the theory of linear electronic networks \cite{control} and the 
theory of plasma waves \cite{plasma}), and the cure for it is well known:
in order to construct an output field that only depends on the past, we have to alter 
the integration contour in the inverse-Fourier transform, going above (with our convention
of Fourier transform) all the poles in the complex plane. 
This procedure, which can be justified rigorously using 
Laplace transforms \cite{laplace}, makes 
the output signal infinitely sensitive to driving forces in the 
infinitely distant past. The reason is simple and well known 
in other contexts: our optical mechanical system possesses instabilities, 
which can be deduced from the homogeneous solution $b_i^{\rm hom}$ 
of Eqs.~(\ref{2.11}), (\ref{2.15}) and (\ref{2.19}), which has 
eigenfrequencies given by $M=0$.
Because the zeros of the equation $M=0$ are generically complex 
and may have positive imaginary parts \cite{BC300}, we end up  
with homogeneous solutions that grow exponentially.
\footnote{~Quadrature operators with \emph{complex} frequency can be defined
by analytical continuations of quadrature operators with \emph{real} 
frequency considered as analytical functions of $\Omega$.}

To quench the instabilities of a SR interferometer 
we have to introduce a proper control system. 
In \cite{BC300} we have given an example of such a control system,  
which we briefly illustrate here. Let us suppose that the 
observed output is $b_{\zeta}$ and we feed back a linear transformation of 
it to control the dynamics of the end mirrors. This operation 
corresponds to making the following substitution in Eq.~(\ref{2.26}):
\beq
\label{2.27}
h \rightarrow h+{\cal C}\,b_{\zeta},
\eeq
where ${\cal C}$ is some \emph{retarded} kernel. Solving again for $b_{\zeta}$, we get:
\bea
b_\zeta^{\cal C}
&&=\frac{1}{{M}_{\cal C}}\,\left [
e^{2 i \beta}
(C_{11}\,\sin\zeta+C_{21}\,\cos\zeta)\,a_1+
e^{2 i \beta}
(C_{12}\,\sin\zeta+C_{22}\,\cos\zeta)\,a_2 \right . \nonumber \\ 
&&\left .  + 
\sqrt{2{\cal K}}\,\tau\,e^{i \beta}\,(D_1\,\sin\zeta+D_2\,\cos\zeta)\,
\frac{h}{h_{\rm SQL}} \right ]\,,
\label{2.28}
\eea
simply replacing the $M$ in (\ref{2.26}) by $M_{\cal C}$, which
depends on $\cal C$.
Note that, by contrast 
with the \emph{uncontrolled} output Eq.~(\ref{2.20}), 
the output field $b_\zeta^{\cal C}$ is no longer a free electric field, 
i.e.\ a quadrature field defined in half open space, satisfying 
the radiative boundary condition. This is due to the fact that 
part of it has been fed back into the arm cavities. Nevertheless,
in the time domain, $b_\zeta^{\cal C}$ commutes with itself at different times. 
In \cite{BC300} we have shown that there 
exists a well-defined $\cal C$ that makes Eq.~(\ref{2.28}) well defined 
in the time domain, getting rid of the instabilities. As a consequence, 
$M_{\cal C}$ has zeros only in the lower-half complex plane and 
we can neglect the homogeneous solution 
$M_{\cal C}\,b_{\zeta}^{{\cal C}\, \rm hom}  =0$
because it decays exponentially in time.

Finally, let us remember the important fact that the introduction of 
this kind of  control system only changes the normalization of the output field.
As a consequence, the noise spectral density is \emph{not} affected. 
However, an extra noise will be present 
due to the electronic device that provides the control 
force on the end mirrors. K. Strain estimated 
that it can be kept smaller than about $10 \%$ of the quantum noise \cite{KS}.

\section{Features of noise spectral density in SR interferometers}
\label{sec3}
In light of the discussion at the end of the last section, we 
shall use Eq.~(\ref{2.28}) as the starting point of
our derivation of the noise spectral density of a (stabilized) SR 
interferometer.

\subsection{Evaluation of the noise spectral density: 
going below the standard quantum limit}
\label{subsec3.1}
The noise spectral density is calculated as follows~\cite{KLMTV00}.
Equation (\ref{2.28}) tells us that the interferometer noise, 
expressed as an equivalent gravitational-wave Fourier 
component, is 
\beq
h_n \equiv  \frac{h_{\rm SQL}}{\sqrt{2{\cal K}}}\,\Delta b_{\zeta}\,,
\label{3.1}
\eeq
where
\beq
\Delta b_{\zeta} = \frac{(C_{11}\,\sin\zeta+C_{21}\,\cos\zeta)\,a_1+
(C_{12}\,\sin\zeta+C_{22}\,\cos\zeta)\,a_2}
{\tau\,(D_1\,\sin\zeta+D_2\,\cos\zeta)}\,.
\label{3.2}
\eeq
Then the (single-sided) spectral density $S^{\zeta}_h(f)$, with $f = \Omega/2\pi$, 
associated with the noise $h_n$ can be computed by the formula (Eq.~(22) 
of Ref. \cite{KLMTV00}) 
\beq
\frac{1}{2}\,2\pi\,\delta(\Omega - \Omega^\prime)\,S^\zeta_h(f) 
= \langle {\rm in} | h_n(\Omega)\,h_n^\dagger(\Omega^\prime) 
|{\rm in} \rangle_{\rm sym}
\equiv \frac{1}{2}
\langle {\rm in} | h_n(\Omega)\,h_n^\dagger(\Omega^\prime) + 
h_n^\dagger(\Omega^\prime)\,h_n(\Omega)|{\rm in} \rangle \,.
\label{3.3}
\eeq
Here we put the superscript $\zeta$ on $S^{\zeta}_h$ to remind 
ourselves that this is the noise when the output is monitored 
at carrier phase $\zeta$ by homodyne detection. 
Assuming that the input of the whole SR interferometer 
is in its vacuum state, as is planned for LIGO-II, i.e.\ 
$|{\rm in} \rangle = |0_a \rangle$, and using  
\beq
 \langle 0_a| a_i\,a^\dagger_{j^\prime} |0_a \rangle_{\rm sym} = 
\frac{1}{2}\,2\pi\,\delta(\Omega- \Omega^\prime)\,
\delta_{i j}\,,
\label{3.4}
\eeq 
(Eq.~(25) of Ref.~\cite{KLMTV00})
we find that Eq.~(\ref{3.3}) can be recast in the simple form (note that $C_{ij}\in\Re$):
\beq
\label{3.5}
S_h^{\zeta}= \frac{h_{\rm SQL}^2}{2{\cal K}}\,
\frac{\left(C_{11}\,\sin\zeta+C_{21}\,\cos\zeta\right)^2+
\left(C_{12}\,\sin\zeta+C_{22}\,\cos\zeta\right)^2}
{\tau^2\,\left|D_1\,\sin\zeta+D_2\,\cos\zeta\right|^2}\,.
\eeq

For comparison, 
let us recall some properties of the noise spectral density 
for conventional interferometers  
(for a complete discussion see Ref.~\cite{KLMTV00}). 
To recover this case we have to take the limit 
$\phi \rightarrow 0$ and $\rho \rightarrow 0$ in the above equations 
or simply use Eq.~(\ref{2.11}) (in a conventional 
interferometer there are no instabilities). In particular, 
for a conventional interferometer,  
Eqs.~(\ref{2.26}) and (\ref{3.1}) take the much simpler form
\footnote{~Note that our definition of $\zeta$ differs from 
the one used in \cite{KLMTV00}} 
\beq
b^{\rm conv}_\zeta = \cos \zeta \, \{ \left [a_2 + (\tan \zeta - {\cal K})\,a_1
\right ]\,e^{2i\beta} \}\,,\quad \quad 
h_n^{\rm conv} = \frac{h_{\rm SQL}}{\sqrt{\cal K}}\,e^{i \beta}\,
\left [a_2 + (\tan \zeta -{\cal K})\,a_1 \right ]\,,
\label{3.6}
\eeq
and the noise spectral density reads
\beq
S_h^{\zeta, \rm conv} 
= \frac{h_{\rm SQL}^2}{2 {\cal K}}\,\left [ 1 + (\tan \zeta - {\cal K})^2 \right ]\,.
\label{3.7}
\eeq
As has been much discussed by Matsko, Vyatchanin and Zubova \cite{HFD} and 
by KLMTV \cite{KLMTV00}, and as 
we shall see in more detail in Sec.~\ref{subsec3.2}, taking as the output 
$b_\zeta$, instead of the quadrature $b_2$ in which 
all the signal $h$ is encoded, builds up correlations between  
shot noise and radiation-pressure noise. We refer 
to correlations of this kind, which are introduced by the 
special read-out scheme, as
{\em static} correlations by contrast with those produced 
by the SR mirror, which we call {\em dynamical} since they are built up
dynamically, as we shall discuss in Sec.~\ref{sec4}. 
The static correlations allow the noise curves for 
a conventional interferometer to go below the SQL when $I_o = I_{\rm SQL}$, 
as was originally observed by 
Matsko, Vyatchanin and Zubova \cite{HFD}. However, if
$\zeta$ is frequency independent as it must be when one 
uses conventional homodyne detection, then the SQL is beaten, 
$S_h^{\zeta, \rm conv} \leq h^2_{\rm SQL}$, only  
over a rather narrow frequency band and only by a very modest amount.
On the other hand, as Matsko, Vyatchanin and Zubova \cite{HFD}
showed, and one can see from Eq.~(\ref{3.7}),  
if we could make the homodyne detection angle $\zeta$ frequency dependent, then 
choosing~\cite{KLMTV00} $\zeta(\Omega) \equiv \arctan {\cal K}(\Omega)$,
would remove completely (in the absence of optical losses) 
the second term in the square parenthesis of Eq.~(\ref{3.7}), 
which is the radiation-pressure noise, leaving only the shot noise 
in the interferometer output, i.e.\ $S_h^{\zeta, \rm conv} 
= {h_{\rm SQL}^2}/{2 {\cal K}}$. 
In order to implement frequency dependent homodyne detection, 
KLMTV \cite{KLMTV00} have recently propose to place 
two 4km-long filter cavities at the interferometer dark port 
and follow them by conventional homodyne detection.
This experimentally  challenging proposal
would allow the interferometer to beat the SQL at frequency 
$f = 100$ Hz by a factor 
$\sqrt{S_h^{\rm conv}}/\sqrt{S_h^{\rm SQL}} \! \sim \! 0.24$, 
over a band of $\Delta f \!\sim\! f$,  at light power $I_o=I_{\rm SQL}$,
and by $\sqrt{S_h^{\rm conv}}/\sqrt{S_h^{\rm SQL}} \! \sim \! 0.18$ if $I_o 
\simeq 3.2 I_{\rm SQL}$.
In conclusion, already in conventional interferometers 
it is possible to beat the SQL provided that we measure 
$b_\zeta$ and build up proper {\em static} correlations 
between shot noise and radiation-pressure noise.
\begin{figure}
\centerline{
\epsfig{file=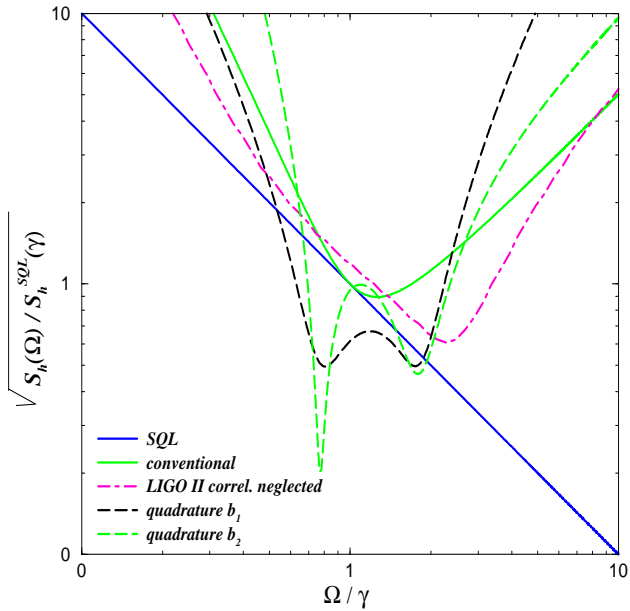,width=0.5\textwidth,height = 0.5\textwidth,angle=-90}}
\vskip 0.1truecm
\caption
{Log-log plot of $\sqrt{S_h(\Omega)}/\sqrt{S_h^{\rm SQL}(\gamma)}$ 
versus $\Omega/\gamma$ for (i) the quadratures $b_1^{\cal C}$ 
($\zeta = \pi/2$) and $b_2^{\cal C}$ ($\zeta = 0$) with
$\rho = 0.9$, $\phi = \pi/2 -0.47$ and $I_o = I_{\rm SQL}$,
(ii) the SQL, (iii) a conventional interferometer with $I_o=I_{\rm SQL}$,
and (iv) the noise curve of LIGO-II \protect\cite{GSSW99}
one would obtain if shot-noise / radiation-pressure correlations were 
(naively) neglected.
For LIGO-II, $\gamma =2\pi \times 100\,{\rm Hz}$ (top axis) and 
$\sqrt{S_h^{\rm SQL}(\gamma)} = 2\times 10^{-24}\,{\rm Hz}^{-1/2}$. 
These curves do not include seismic and thermal noises; 
for LIGO-II the latter is expected to be 
slightly above the SQL \protect\cite{BGV00}.} 
\label{Fig2}
\end{figure}

Let us now go back to SR interferometers. They have 
the interesting property of building up {\it dynamically} the 
correlations between shot noise and radiation-pressure noise,  
thanks to the SR mirror. Indeed, even if we restrict ourselves 
to the noise curves associated with the two quadratures $b_1^{\cal C}$ 
and $b_2^{\cal C}$, i.e.\ we 
do not measure $b_\zeta^{\cal C}$, the SR interferometer can still 
go below the SQL. 
Moreover, if the SR interferometer works at the SQL power, i.e.\ $I_o = I_{\rm SQL}$, 
as is tentatively planned for LIGO-II, then the noise curves [Eq.~(\ref{3.5})]  
can exhibit one or two resonant dips whose depths increase and widths 
decrease as the SR-mirror's reflectivity is raised. 
(We postpone the discussion of this interesting feature to Sec.~\ref{sec4}.)
These resonances allow us to reshape the noise curves  and 
beat the SQL by much larger amounts than in a conventional 
interferometer with static correlations introduced by frequency-independent 
homodyne detection.

More specifically,  the noise spectral density, Eq.~(\ref{3.5}), depends on the 
physical parameters which characterize the SR interferometer (see Table~\ref{Tab1}): 
the light power $I_o$, the SR detuning $\phi$, the reflectivity 
of the SR mirror $\rho$ and the homodyne phase $\zeta$. 
To give an example of  LIGO-II noise curves,
in Fig.~\ref{Fig2} we plot the $\sqrt{S_h(\Omega)}$ for the two 
quadratures $b_1^{\cal C}$ ($\zeta = \pi/2$) and $b_2^{\cal C}$ ($\zeta =0$), 
for: $\rho = 0.9$, $\phi = \pi/2 -0.47$ and $I_o = I_{\rm SQL}$.
Also shown for comparison are the SQL line, the noise 
curve one would obtain if one ignored the correlations between 
the shot noise and radiation-pressure noise\cite{GSSW99}, \footnote{~
Before the research reported in this paper, the LIGO 
community computed the noise curves for SR interferometers 
by (i) evaluating the shot noise $S_h^{\rm shot}$, (ii) then 
({\it naively} 
assuming no correlations between shot noise 
and radiation-pressure noise) using the uncertainty principle 
$S_h^{\rm shot}\,S_h^{\rm RP} \geq \left(S_h^{\rm SQL}\right)^2/4$, with the 
equality sign to evaluate the radiation-pressure noise $S_h^{\rm RP}$, (iii) then 
adding the two. This procedure gave the noise curve labeled 
``correlations neglected'' in Fig.~\ref{Fig2}; see Fig.~2 of Ref.~\cite{GSSW99}.}
and for a 
conventional interferometer with $I_o=I_{\rm SQL}$ and $\zeta =0$, explicitly 
given by \cite{KLMTV00}: 
\beq
S_{h}^{\zeta =0, \rm conv} = \frac{S_h^{\rm SQL}}{2}\,\left ( {\cal K} + 
\frac{1}{\cal K} \right )\,.
\label{3.8}
\eeq
\begin{figure}
\centerline{
\epsfig{file=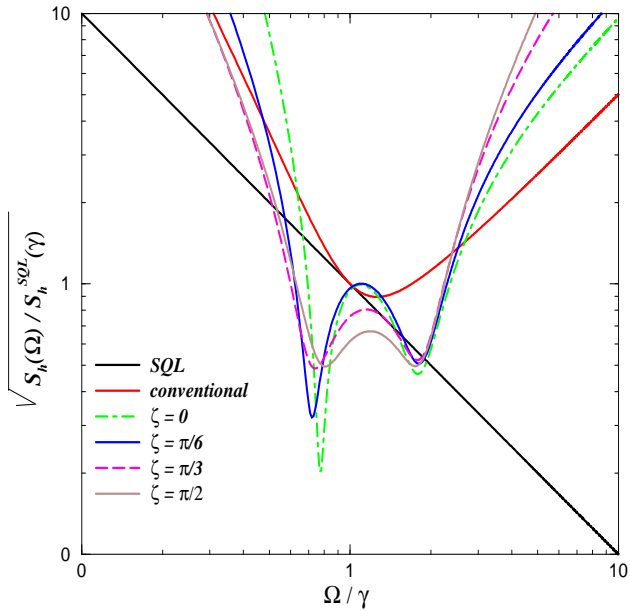,width=0.5\textwidth,height = 0.5\textwidth,angle=-90}}
\vskip 0.2truecm
\caption{Log-log plot of $\sqrt{S_h(\Omega)/S^{\rm SQL}_h(\gamma)}$ 
versus $\Omega/\gamma$ for the following choices of the 
frequency independent homodyne phase: $\zeta = 0$, $\zeta = \pi/6$, 
$\zeta = \pi/3$  and $\zeta =\pi/2$, with $\rho = 0.9$, 
$\phi = \pi/2 -0.47$ and $I_o = I_{\rm SQL}$. 
The plot also shows the noise curve for a conventional interferometer 
and the SQL line. For LIGO-II, $\gamma =2\pi \times 100\,{\rm Hz}$ (top axis) and 
$\sqrt{S_h^{\rm SQL}(\gamma)} = 2\times 10^{-24}\,{\rm Hz}^{-1/2}$. } 
\label{Fig3}
\end{figure}
The sensitivity curves for the two quadratures 
go substantially below the SQL and show two interesting 
resonant valleys. In Fig.~\ref{Fig3} we plot the noise curves  
$\sqrt{S_h(\Omega)}$ for different values of the frequency independent homodyne 
angle $\zeta$, choosing 
the same parameters used in Fig.~\ref{Fig2}, i.e.\ $\rho = 0.9$, $\phi = \pi/2 -0.47$ 
and $I_o = I_{\rm SQL}$. Note that the location of the resonant dips does 
not depend much on the angle $\zeta$. This property is confirmed analytically 
in Sec.~\ref{sec4} in the case of a highly-reflecting 
SR mirror, by an analysis that elucidate the underlying physics.

Before ending this section, let us give an idea of the performances
achievable in a SR interferometer 
{\it if its thermal noise can be made negligible} \cite{BGV00}. 
We have estimated the signal-to-noise ratio for inspiraling binaries, which 
are among the most promising sources for the detection of GW with earth-based interferometers.
The square of the signal-to-noise ratio for a binary system made of black holes 
and/or neutron stars is given by:
\beq
\left (\frac{S}{N} \right )^2 =4 
\int_{0}^{+\infty} \frac{|h(f)|^2}{S_h(f)}\,df\,.
\label{3.9}
\eeq
Using the Newtonian quadrupole approximation, for which 
the waveform's Fourier transform is $|h(f)|^2 \propto f^{-7/3}$,
and introducing in the above integral a lower cutoff due to 
seismic noise at $\Omega_s =0.1\gamma$ ($f_s \simeq 10$ Hz), we get 
for the parameters used in Fig.~\ref{Fig2}: 
\beq
\frac{({S}/{N})_{1}}{({S}/{N})_{\rm conv}} \simeq 1.83\,,\quad \quad
\frac{({S}/{N})_{2}}{({S}/{N})_{\rm conv}} \simeq 1.98\,,
\label{3.10}
\eeq
where $({S}/{N})_{1}$, $({S}/{N})_{2}$ and 
$({S}/{N})_{\rm conv}$ use for the noise spectral density 
either that of the first quadrature $b_1^{\cal C}$ or the second 
quadrature $b_2^{\cal C}$ or the conventional 
interferometer, respectively.
A more thorough analysis of signal-to-noise 
ratio for inspiraling binaries inevitably requires the specification 
of the readout scheme and we plan to publish it elsewhere \cite{BC400}.

\subsection{Effective shot noise and radiation-pressure noise}
\label{subsec3.2}
In this section we shall discuss the crucial role 
played by shot-noise\,/\,radiation-pressure correlations that
are present in LIGO-II's quadrature outputs (\ref{2.20}) and 
noise spectral densities (\ref{3.5}), 
in beating the SQL. Our analysis is based on the general formulation 
of linear quantum measurement theory developed by Braginsky 
and Khalili~\cite{BK92} 
and assumes also the results obtained in \cite{BGKMTV00}, \cite{BC300}. 

To identify the radiation pressure and the shot noise 
contributions in the total optical noise, 
we use the fact that they transform  
differently under rescaling of the mirror mass. Indeed, it is straightforward 
to show that in the total optical noise there exist 
only two kinds of terms. There are terms that are invariant under  
rescaling of the mass and terms 
that are proportional to $1/m$. Hence, quite generally 
we can rewrite the output ${\cal O}$ of the 
whole optical system as \cite{BK92}, \cite{BC300}
\beq
{\cal O}(\Omega) = {\cal Z}(\Omega) + {\cal R}_{xx}(\Omega)\,{\cal F}(\Omega) + 
L\,h(\Omega)\,,
\label{3.11}
\eeq
where by output we mean one of the two quadratures
$b_1^{\cal C}$, $b_2^{\cal C}$ or a combination of them, 
e.g., $b_\zeta^{\cal C}$ (modulo a normalization factor)
and where ${\cal R}_{xx}$ is the susceptibility of the antisymmetric mode 
of motion of the four mirrors~\cite{BK92}, given by
\beq
{\cal R}_{xx}(\Omega)=-\frac{4}{m\,\Omega^2}\,.
\label{3.12}
\eeq
The observables ${\cal Z}$ and ${\cal F}$ in Eq.~(\ref{3.11})
do not depend on the mirror masses $m$, and satisfy the 
commutation relations (Eq.~(2.19) in Ref.~\cite{BC300})
\beq
[{\cal F}(\Omega),{\cal F}^\dagger(\Omega^\prime)] = 0=
[{\cal Z}(\Omega),{\cal Z}^\dagger(\Omega^\prime)]\,, \quad \quad 
[{\cal Z}(\Omega),{\cal F}^\dagger(\Omega^\prime)] = 
-2 \pi i \hbar\delta(\Omega - \Omega^\prime)\,.
\label{3.15}
\eeq
We shall refer to ${\cal Z}$ and ${\cal F}$ as the 
{\em effective} shot noise and {\em effective} radiation-pressure force, 
respectively,
because we have shown \cite{BC300} that 
for a SR interferometer the {\it real} back-action force
acting on the test masses is not proportional to
the effective radiation-pressure noise, 
but instead is a combination of the two effective observables 
${\cal Z}$ and ${\cal F}$. 
When the shot noise and radiation-pressure noise are correlated, the real 
back-action force does not commute with itself at different times,  
\footnote{~We have shown \cite{BC300} that as a consequence of this  
identification the antisymmetric mode of motion of the four mirrors 
acquires an optical-mechanical rigidity and a SR interferometer responds to 
GW signal like an optical spring. This phenomenon 
was already observed in optical bar detectors by  
Braginsky's group \cite{OB}.}
which makes the analysis in terms of real quantities more complicated
than in terms of the effective ones. We prefer to discuss our results 
in terms of real quantities separately \cite{BC300}, in a more 
formal context which uses the description of a GW interferometer as a  
linear quantum-measurement device \cite{BK92}.

The noise spectral density, written in terms 
of the effective operators ${\cal Z}$ and ${\cal F}$, reads~\cite{BK92}: 
\beq
\label{3.13}
S_{h}=\frac{1}{L^2}\,\left\{
       S_{{\cal Z} {\cal Z}}
      +2{\cal R}_{xx}\,\Re\left[S_{{\cal F} {\cal Z}}\right] 
+ {\cal R}_{xx}^2\, S_{{\cal F} {\cal F}}\right\}\,, 
\eeq
where the (one-sided) cross spectral density of two operators 
is expressible, by analogy with Eq.~(\ref{3.3}), as
\beq
\label{3.14}
\frac{1}{2}\, S_{{\cal A} {\cal B}}(\Omega) 
2\pi\,\delta\left(\Omega-\Omega'\right)= 
\frac{1}{2}\,\langle {\cal A}(\Omega) {\cal B}^\dagger(\Omega')+ {\cal B}^\dagger
(\Omega') {\cal A}(\Omega)\rangle\,.
\eeq
In Eq.~(\ref{3.13}) the terms containing $S_{{\cal Z} {\cal Z}}$, $S_{{\cal F} 
{\cal F}}$ and $\Re\left[S_{{\cal F} {\cal Z}}\right]$ 
should be identified as effective shot noise, back-action noise and a term 
proportional to the effective correlation between the two noises,
respectively \cite{BK92}.
{}Relying on the commutators (\ref{3.15}) between the effective field operators 
one can derive \cite{BK92}, \cite{BC300} the following  
uncertainty relation for the (one-sided) spectral densities and cross 
correlations of ${\cal Z}$ and ${\cal F}$:
\beq
\label{3.16}
S_{{\cal Z} {\cal Z}}\,S_{{\cal F} {\cal F}} - 
S_{{\cal Z} {\cal F}}\,S_{{\cal F} {\cal Z}} \ge \hbar^2\,.
\eeq
Equation ~(\ref{3.16}) does not, in general, impose a lower bound on the noise spectral
density Eq.~(\ref{3.13}). 
However, in a very important type of measurement it does, namely   
for interferometers with uncorrelated shot noise and back-action 
noise, e.g., LIGO-I/TAMA/Virgo.
In this case $S_{{\cal Z}{\cal F}}=0=S_{{\cal F}{\cal Z}}$ \cite{KLMTV00}  
and inserting the vanishing correlations into Eqs.~(\ref{3.14}), 
(\ref{3.16}), one easily finds that the noise spectral density has a lower 
bound which is given by  the standard quantum limit, i.e.
\beq
S_{h}(\Omega)\ge S_h^{\rm SQL}(\Omega)\equiv
\frac{2 \left|R_{xx}(\Omega)\right| \hbar}{L^2} = \frac{8 \hbar}{m\Omega^2\,L^2}
=h_{\rm SQL}^2(\Omega)\,.
\label{3.17}
\eeq
{}From this it follows that to beat the SQL one must create
correlations between shot noise and back-action noise. 

Before investigating those correlations in a SR inteferometer, 
we shall first show 
how such  correlations can be built up statically in a conventional  
(LIGO-I/TAMA/Virgo) interferometer by implementing 
frequency-independent homodyne detection at some angle $\zeta$
~\cite{HFD}, \cite{KLMTV00}. 
By identifying in the interferometer output (\ref{3.6})  
the terms independent of $m$  as effective shot noise  
and those inversely proportional to $m$
as effective back-action noise, we get the effective field operators 
${\cal Z}^{\rm conv}_\zeta$ and ${\cal F}^{\rm conv}_\zeta$:
\beq
{\cal Z}^{\rm conv}_\zeta(\Omega) = \frac{e^{i\beta}\,L\,h_{\rm SQL}}{\sqrt{2{\cal K}}}\,
\left( a_2 + a_1 \,\tan \zeta  \right)\,, 
\quad \quad 
{\cal F}^{\rm conv}_\zeta(\Omega) = 
\frac{\hbar\,e^{i\beta}\,\sqrt{2{\cal K}}}{L\,h_{\rm SQL}}\,a_1\,.
\label{3.18}
\eeq
[We remind the readers that $h_{\rm SQL}\propto 1/\sqrt{m}$ and that ${\cal K}\propto 1/m$.]
Evaluating the spectral densities of those operators 
using Eqs.~(\ref{3.14}) and (\ref{3.4}), we obtain the following expressions for the 
spectral densities and their static correlations:
\beq
S^{\rm conv}_{{\cal Z}_\zeta {\cal Z}_\zeta}(\Omega) = \frac{L^2\,h_{\rm SQL}^2}{2{\cal K}}
\,(1 + \tan^2 \zeta)
\,, \quad \quad 
S^{\rm conv}_{{\cal F}_\zeta {\cal F}_\zeta}(\Omega) = \frac{2{\cal K}\,\hbar^2}
{L^2\,h_{\rm SQL}^2}\,, 
\quad \quad S^{\rm conv}_{{\cal Z}_\zeta {\cal F}_\zeta} (\Omega) = \hbar \,\tan \zeta=
S^{\rm conv}_{{\cal F}_\zeta {\cal Z}_\zeta}(\Omega)\,. 
\label{3.19}
\eeq
By inserting these in Eq.~(\ref{3.13}) and optimizing the coupling constant 
${\cal K}$, we see that the SQL can be beaten for any $0<\zeta< \pi/2$, i.e.\ 
whenever there are nonvanishing correlations. See Refs. \cite{KLMTV00} and \cite{HFD} 
for further details. 

Let us now derive the correlations between shot noise and back action 
noise in SR interferometers. Because in this case 
the correlations are built up dynamically by the SR mirror 
and are present in all quadratures, as an example, we limit ourselves 
to the two quadratures $b^{\cal C}_1$ and $b^{\cal C}_2$. 
Identifying in Eqs.~(\ref{3.1}), (\ref{3.2})
the effective shot and back-action noise terms due to their $m$ dependences, 
we obtain the effective field operators 
${\cal Z}_1$, ${\cal Z}_2$, ${\cal F}_1$ and ${\cal F}_2$
\bea
&& {\cal Z}_{1}(\Omega)= -\frac{e^{i\beta}\,L\,h_{\rm SQL}}{\sqrt{2{\cal K}}}\,
\frac{\left [a_1\,(-2\rho\,\cos 2\beta + (1+\rho^2)\,\cos 2\phi) + 
a_2\,(-1+\rho^2)\,\sin 2\phi \right ]\,\csc \phi}{\tau\,(1 + e^{2i\beta}\,\rho)}\,,
\nonumber \\
&& {\cal Z}_{2}(\Omega)= -\frac{e^{i\beta}\,L\,h_{\rm SQL}}{\sqrt{2{\cal K}}}\,
\frac{\left [ a_1\,(1-\rho^2)\,\sin 2\phi + 
a_2\,(-2\rho\,\cos 2\beta + (1+\rho^2)\,\cos 2\phi)\right ]\,\sec 
\phi }{\tau\,(-1 + e^{2i\beta}\,\rho)}\,,
\label{3.20}
\eea
and 
\bea
&& {\cal F}_{1}(\Omega)= \frac{\hbar\,e^{i\beta}\,\sqrt{2{\cal K}}}{L\,h_{\rm SQL}}\,
\frac{\left [ a_1\,(1+\rho^2)\,\cos \phi + 
a_2\,(-1+\rho^2)\,\sin \phi \right ]}{\tau\,(1 + e^{2i\beta}\,\rho)}\,,
\nonumber \\
&& {\cal F}_{2}(\Omega)= \frac{\hbar\,e^{i\beta}\,\sqrt{2{\cal K}}}{L\,h_{\rm SQL}}\,
\frac{\left [ a_1\,(-1+\rho^2)\,\cos \phi + 
a_2\,(1+\rho^2)\,\sin \phi \right ]}{\tau\,(-1 + e^{2i\beta}\,\rho)}\,.
\label{3.21}
\eea  
Evaluating the spectral densities of the above operators 
through Eqs.~(\ref{3.14}) and (\ref{3.4}) we obtain the following expressions:
\bea
&& S_{{\cal F}_{1}{\cal F}_{1}}(\Omega) = 
\frac{\hbar^2\,2{\cal K}}{L^2\,h_{\rm SQL}^2}\,\frac{1 + \rho^4 +2\rho^2\cos \phi}{
{(1-\rho^2)\,(1 + \rho^2 +2\rho\,\cos 2 \beta)}}\,,\nonumber \\
&& S_{{\cal F}_{2}{\cal F}_{2}}(\Omega) = 
\frac{\hbar^2\,2{\cal K}}{L^2\,h_{\rm SQL}^2}\,\frac{1 + \rho^4 -2\rho^2\cos \phi}{
{(1-\rho^2)\,(1 + \rho^2 -2\rho\,\cos 2 \beta)}}\,,
\label{3.22}
\eea
and 
\bea
&& S_{{\cal Z}_{1}{\cal Z}_{1}}(\Omega) = 
\frac{L^2\,h_{\rm SQL}^2}{2{\cal K}}
\frac{\left [ 4(-1+\rho^2)^2\,\cos^2 \phi + (-2\rho\,\cos 2\beta + (1+\rho^2)\,
\cos 2\phi)^2\,\csc^2 \phi \right]}
{(1-\rho^2)\,(1 + \rho^2 +2\rho\,\cos 2 \beta)}\,, \nonumber \\
&& S_{{\cal Z}_{2}{\cal Z}_{2}}(\Omega) = 
\frac{L^2\,h_{\rm SQL}^2}{2{\cal K}}
\frac{\left [ 4(-1+\rho^2)^2\,\sin^2 \phi + (-2\rho\,\cos 2\beta + (1+\rho^2)\,
\cos 2\phi)^2\,\sec^2 \phi \right]}
{(1-\rho^2)\,(1 + \rho^2 -2\rho\,\cos 2 \beta)}\,.
\label{3.23}
\eea
Finally, for the correlations between the shot noise and back-action noise we get:
\bea
&& S_{{\cal F}_{1}{\cal Z}_{1}}(\Omega)
=S_{{\cal Z}_{1}{\cal F}_{1}}(\Omega)=
-\frac{\hbar\,\left [ (-1 + \rho^2)^2 -2\rho\,(1 + \rho^2)\,\cos 2\beta + 4\rho^2\,\cos 2\phi \right ]\,
\cot \phi}{(1-\rho^2)\,(1 + \rho^2 + 2\rho\,\cos 2 \beta)} \,, \nonumber \\
&& S_{{\cal F}_{2}{\cal Z}_{2}}(\Omega)=S_{{\cal Z}_{2}{\cal F}_{2}}(\Omega)=
\frac{\hbar\,\left [ (-1 + \rho^2)^2 +2\rho\,(1 + \rho^2)\,\cos 2\beta - 4\rho^2\,\cos 2\phi \right ]\,
\tan\phi}{(1-\rho^2)\,(1 + \rho^2 - 2\rho\,\cos 2 \beta)}\,.
\label{3.24}
\eea
These correlations depend on the sideband angular frequency $\Omega$ and 
are generically different from zero. However, when 
$\phi=0$ and $\phi=\pi/2$ the correlations are zero.  
We shall analyze these two extreme configurations in the 
following section. 

\subsection{Two special cases: extreme signal-recycling and 
resonant-sideband-extraction configurations}
\label{subsec3.3}

In this section we discuss two extreme cases that are well known 
and have been much investigated 
in the literature using a semiclassical analysis \cite{SR,M95}. 
In these two cases the dynamical correlations between 
shot noise and radiation-pressure noise are zero. This 
has two implications: (i) the semiclassical analysis and 
predictions \cite{SR,M95} are correct (when straightforwardly 
complemented by radiation pressure noise), and (ii)  
the noise curves are always above the SQL.
Of course, static correlations can always be introduced 
by measuring the quadrature $b_\zeta$. In these two extreme 
cases there are no instabilities and 
the input--output relation of the SR interferometer 
can be obtained from the conventional 
noise by just rescaling the parameter ${\cal K}$ [Eq.~(\ref{2.13})].

\subsubsection{Extreme signal-recycling (ESR) configuration:  $\phi=0$}
\label{subsubsec3.3.1}
\begin{figure}
\begin{center}
\begin{tabular}{cc}   
\hspace{-0.8cm} 
\epsfig{file=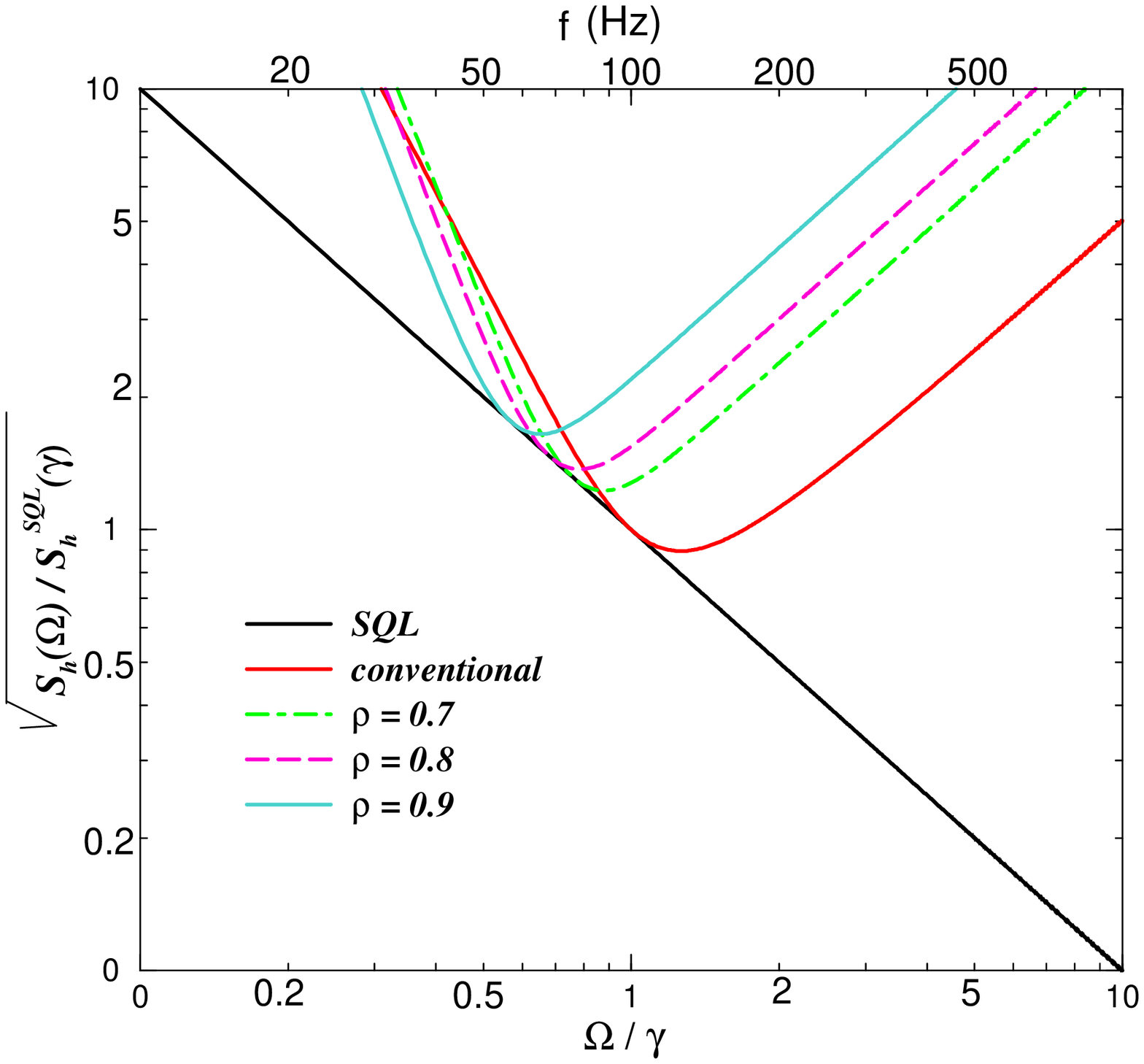,width=0.45\textwidth,height = 0.45\textwidth,angle=-90} &
\hspace{1cm}
\epsfig{file=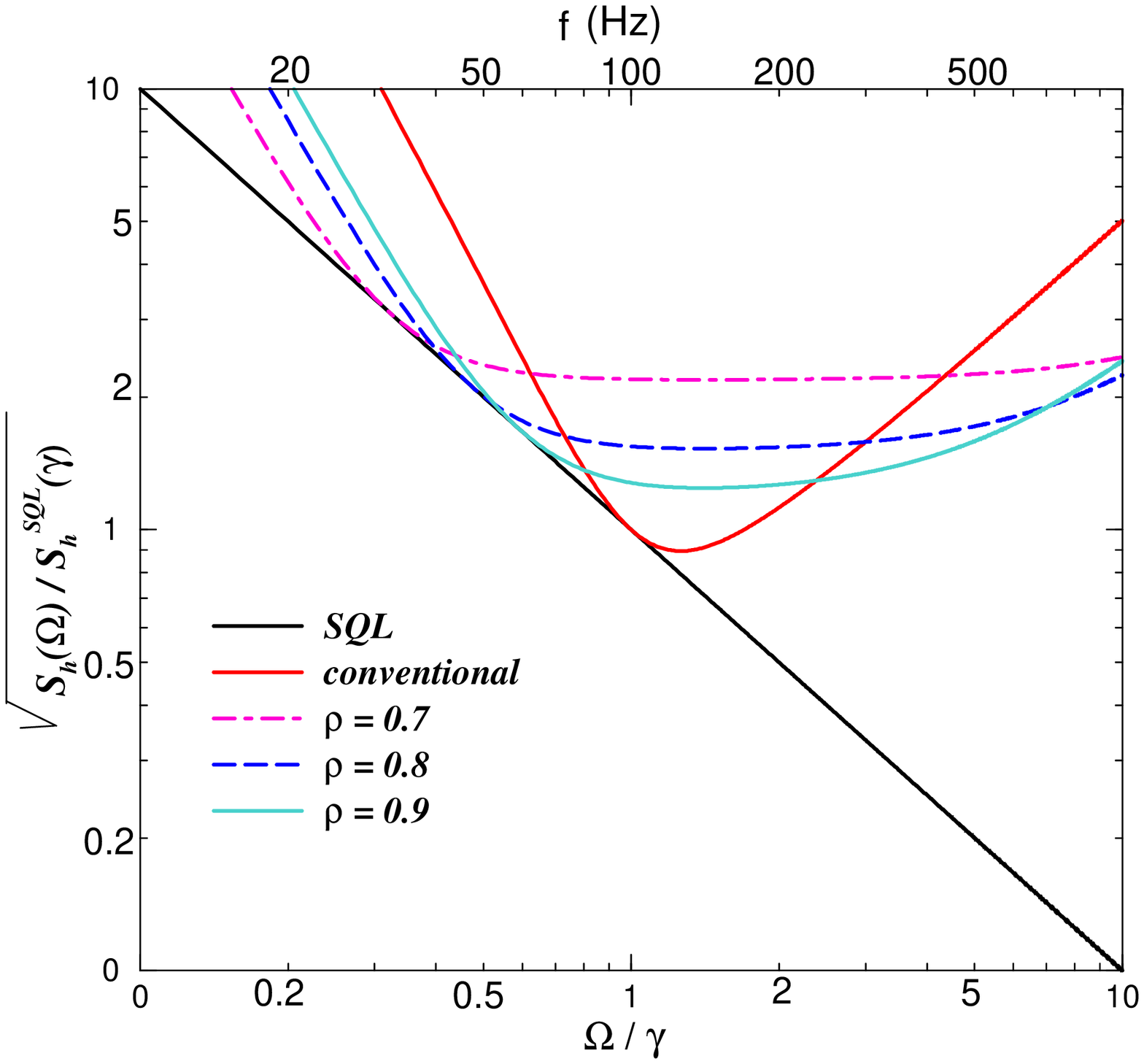,width=0.45\textwidth,height = 0.45\textwidth,angle=-90}
\end{tabular}
\caption{\sl Log-log plot of $\sqrt{S_h^{\rm ESR}(\Omega)/S_h^{\rm SQL}(\gamma)}$ 
versus $\Omega/\gamma$ for the extreme signal-recycling 
configuration (left panel) $\phi=0$ with $\rho = 0.7$, $\rho = 0.8$,  
$\rho = 0.9$ and $I_o = I_{\rm SQL}$ and 
for the extreme resonant-sideband-extraction configuration 
(right panel) $\phi = \pi/2$ with 
$\rho = 0.7$, $\rho = 0.8$ and  $\rho = 0.9$,  
with $I_o = I_{\rm SQL}$. Also plotted for comparison are the noise curve for a 
conventional interferometer and the SQL line. For further detail on these 
well known configurations, see Refs. \protect\cite{SR,M95}.}
\label{Fig4}
\end{center}
\end{figure}

For $\phi=0$, the gravitational-wave signal appears only in the second 
quadrature $b_2$ but not in the first quadrature $b_1$ 
(see Eq.~(\ref{2.26}) with $\zeta =0$ and $\pi/2$, respectively). Defining
\beq
\label{3.25}
\widetilde{{\cal K}}\equiv\frac{{\cal K}\,\tau^2}{1+\rho^2 -2\rho\,\cos{2\beta}}\,,
\eeq
it is straightforward to deduce that the spectral density of the noise 
takes the simple form
\beq
\label{3.26}
S^{\rm ESR}_h
= \frac{S_h^{\rm SQL}}{2}\, \left ( \frac{1}{\widetilde{{\cal K}}} + 
\widetilde{{\cal K}} \right )\,.
\eeq
In the left panel of Fig.~\ref{Fig4} we plot $\sqrt{S_h^{\rm ESR} 
(\Omega)/S_h^{\rm SQL}(\gamma)}$ versus $\Omega/\gamma$ 
for different choices of the reflectivity $\rho$. 
As we vary the reflectivity of the SR mirror 
the minimum of the various curves is shifted along the SQL line, 
and the shape of the noise curve change a bit because both 
${\cal K}$ and $\beta$ in Eqs.~(\ref{3.25}), (\ref{3.26}) 
depend on frequency.  
Moreover, for $\Omega/\gamma \gg 1$ and $\Omega/\gamma \ll 1$ 
the curves are well above the conventional interferometer noise. This effect 
becomes worse and worse as $\rho \rightarrow 1$ and is described by 
the formulas
\beq
\frac{S^{\rm ESR}_h(\Omega)}{S_h^{\rm SQL}(\gamma)} 
\rightarrow \frac{1}{4}\frac{\Omega^2}{\gamma^2}\,
\left (\frac{1 + \rho}{1-\rho} \right)\,\frac{I_{\rm SQL}}{I_o}\,, \quad \quad 
\frac{\Omega}{\gamma} \gg 1\,; \quad \quad 
\frac{S^{\rm ESR}_h(\Omega)}{S_h^{\rm SQL}(\gamma)} \rightarrow
\frac{\gamma^4}{\Omega^4}
\,\left (\frac{1 + \rho}{1-\rho} \right)\,\frac{I_o}{I_{\rm SQL}}\,, \quad \quad 
\quad \frac{\Omega}{\gamma} \ll 1\,.
\label{3.27}
\eeq
The signal-to-noise ratio for inspiraling binaries 
is given in this case (for $\rho = 0.9$, $I_o = I_{\rm SQL}$) by:
\beq
\frac{({S}/{N})_{\rm ESR}}{({S}/{N})_{\rm conv}} \simeq 0.73\,.
\label{3.28}
\eeq
Hence, this LIGO-II configuration ($\phi=0$) is not appealing. 
The noise curves could be better than the ones for a conventional interferometer  
in the range $\sim \! 20-60$ Hz, depending on the value of $\rho$,  
but they get worse everywhere else, and 
overall, for any $\rho$  the signal-to-noise ratio for inspiraling 
binaries is lower than in the case of a conventional interferometer.

\subsubsection{Extreme resonant-sideband-extraction (ERSE) 
configuration:  $\phi=\pi/2$}
\label{subsubsec3.3.2}

For $\phi={\pi}/{2}$, using Eq.~(\ref{2.26}) with $\zeta =\pi/2$,  
we find that only the first quadrature $b_1$ contains 
the gravitational-wave signal. Introducing
\beq
\overline{{\cal K}}\equiv\frac{{\cal K}\,\tau^2}{1+\rho^2+ 2\rho\,\cos{2\beta}}\,,
\label{3.29}
\eeq
(which depends on frequency through both ${\cal K}$ and $\beta$),  
we easily deduce that the noise spectral density reads
\beq
S_h^{\rm ERSE} = 
\frac{S_h^{\rm SQL}}{2}\, \left ( \frac{1}{\overline{{\cal K}}} + 
\overline{{\cal K}} 
\right )\,.
\label{3.30}
\eeq
The right panel of 
Fig.~\ref{Fig4} shows $\sqrt{S_h^{\rm ERSE}(\Omega)/S_h^{\rm SQL}(\gamma)}$ 
as a function of $\Omega/\gamma$
for different values of the reflectivity $\rho$. 
As for the ESR configuration discussed above, 
when we vary the reflectivity of the SR mirror 
the minimum of the various curves moves along the SQL line. 
But by contrast with the ESR configuration, for 
$\Omega/\gamma \gg 1$ and $\Omega/\gamma \ll 1$ 
the curves are significantly below the conventional-interferometer noise. This effect 
becomes better and better as $\rho \rightarrow 1$ and is 
described by the asymptotic limits
\beq
\frac{S^{\rm ERSE}_h(\Omega)}{S_h^{\rm SQL}(\gamma)} 
\rightarrow \frac{1}{4}\frac{\Omega^2}{\gamma^2}\,
\left (\frac{1 - \rho}{1+\rho} \right)\,\frac{I_{\rm SQL}}{I_o}\,, \quad \quad 
\frac{\Omega}{\gamma} \gg 1\,;\quad \quad 
\frac{S^{\rm ERSE}_h(\Omega)}{S_h^{\rm SQL}(\gamma)} \rightarrow
\frac{\gamma^4}{\Omega^4}
\,\left (\frac{1 - \rho}{1+\rho} \right)\,\frac{I_o}{I_{\rm SQL}}\,, \quad \quad 
\frac{\Omega}{\gamma} \ll 1\,.
\label{3.31}
\eeq

In conclusion, in the ERSE configuration ($\phi = {\pi}/{2}$), 
the situation is in some sense the reverse of the ESR scheme ($\phi = 0$).
In the former the bandwidths are much larger than in either the ESR  
of the conventional interferometer. However, the more broadband 
curves are obtained at the cost of losing sensitivity in the frequency 
range $\sim \! 70-250$ Hz and this explains why
the maximum signal-to-noise 
ratio for inspiraling binaries,
\beq
\frac{({S}/{N})_{\rm ERSE}}{({S}/{N})_{\rm conv}} 
\simeq 1.096\, \quad \quad {\rm for} \quad \rho = 0.48 \quad 
{\rm and} \quad I_o = I_{\rm SQL} \,.
\label{3.32}
\eeq
is not very different from that of a conventional interferometer.
Finally, let us observe that our two extreme cases 
are linked mathematically by taking $\rho\rightarrow\ -\rho$ (  
$\overline{{\cal K}} \rightarrow \widetilde{{\cal K}}$) and exchanging 
the two quadratures. For much further analysis and detail 
of the ERSE and ESR configurations, see Refs. \cite{SR,M95}.

\section{Structure of resonances and instabilities}
\label{sec4}
We now turn our attention from the well known extreme configurations, for which 
previous analysis gave correct predictions, to the more general 
case $0< \phi < \pi/2$. 
As Figs.~\ref{Fig2}, \ref{Fig3} show, the noise curves for 
a SR interferometer with frequency independent homodyne detection generically 
exhibit resonant features that vary as $I_o$, $\rho$, $\phi$ and $\zeta$ 
are changed.
These resonances are closely related to the optical-mechanical resonances of 
the dynamical system formed by the optical field and the mirrors.
A thorough study of this system must investigate explicitly 
the motion of the mirrors, instead of including it
implicitly in the formulae as we did in this paper.
It can be most clearly worked out using the formalism of linear quantum 
measurements \cite{BK92}, which we have recently extended 
to SR interferometers~\cite{BC300}. In this section, 
we limit our investigation to the resonant structures in the amplitudes 
of the optical fields, and for simplicity
we work in the limit of a totally reflecting SR mirror, 
i.e.\ $\rho =1$. 
This limit provides simple  analytical expressions for the  
resonant frequencies as functions of the SR detuning phase 
$\phi$ and the light power $I_o$.
We shall comment on the general 
case $\rho \neq 1$, which we have tackled at length in Ref.~\cite{BC300}, 
only at the end of this section.

\subsection{Resonances of the closed system: $\rho =1$}
\label{subsec4.1}
We shall investigate the free oscillation modes of the 
whole interferometer when the GW signal is absent 
[$h(\Omega)=0$] and there is no output field ($\rho=1$), 
so the system is closed. We consider the regime of classical electrodynamics, i.e.\ 
we work with the two classical quadrature fields $E_1$ and $E_2$, 
satisfying the same equations of motion as the quantum-field operators 
$c_1$ and $c_2$ (see Fig.~\ref{Fig1}). 
We shall evaluate the stationary modes, notably the eigenmodes  
and eigenvalues of the whole optico-mechanical system made of the 
end mirrors and the signal recycled optical field. 
We achieve this by propagating the in-going 
fields $E_1$ and $E_2$ (entering the beam splitter's dark port)
into the conventional interferometer, along a complete round trip, 
and then through the SR cavity back to the starting point. 
The round-trip propagation leads to the following 
homogeneous equation for the eigenmodes:
\beq
\left[
\left(\matrix{ \cos 2\phi & -\sin 2\phi \cr \sin 2\phi & \cos 2 \phi }\right)
e^{2 i \beta}
\left(\matrix{ 1 & 0\cr -{\cal{K}} & 1}\right)
-{\rm I} \right]
\left(\matrix{ E_1\cr E_2}\right)=0\,,
\label{4.1}
\eeq
which can be simplified into the more interesting form:
\beq
T
\left(\matrix{ e^{2i(\alpha+\beta)}-1 \cr  0 & e^{2i(-\alpha+\beta)}-1 }\right)
T^{-1}
\left(\matrix{ E_1 \cr E_2}\right)=0\,,
\quad \quad 
2\alpha\equiv\arccos\left(\cos 2 \phi+\frac{\cal K}{2} \sin 2 \phi \right)\,,
\label{4.2}
\eeq
where $T$ is a matrix whose precise form is unimportant. 
Note that the definition of the function $\arccos$ ensures that 
$\Re(2\alpha)$ ranges from $0$ to $\pi$.
The free oscillation condition is then given by:
\beq
\cos 2\beta_{\rm res.} = \cos 2\alpha = \cos 2 \phi+\frac{\cal K}{2} \sin 2 \phi\,.
\label{4.3}
\eeq
Solving Eq.~(\ref{4.3}) explicitly in terms of the frequency $\Omega$, 
we obtain the rather simple analytical equation for the position of the resonances:
\beq
\frac{\Omega^2_{\rm res.}}{\gamma^2} = \frac{1}{2}\left [ \tan^2 \phi \pm 
\sqrt{\tan^4 \phi - \frac{4I_o}{I_{\rm SQL}}\,
\tan \phi} \right ]\,.
\label{4.4}
\eeq
This equation is characterized by three regimes ($0<\phi < \pi$):
\begin{itemize}
\item {$\phi > \pi/2$}: one real and one imaginary resonant 
frequency;
\item {$\arctan[(4I_o/I_{\rm SQL})^{1/3}] < \phi < \pi/2$}:
two real resonant frequencies;
\item {$0< \phi < \arctan[(4I_o/I_{\rm SQL})^{1/3}]$}:
two complex conjugate resonant frequencies.
\end{itemize}
Equation (\ref{4.4}) is very similar to the resonance 
equation that Braginsky, Gorodetsky and Khalili have 
derived for their proposal ``Optical bar gravitational wave 
detectors'' (see Appendix D of Ref.~\cite{OB}).

For very low light power, ${I_o} \ll {I_{\rm SQL}}$, 
the second term under the square root on the RHS of Eq.~(\ref{4.4}) goes to $0$ and 
the four roots tend to $\Omega = 0$ (double root) and $\Omega=\pm \gamma \tan\phi$.
We interpret this limit as follows (see Ref.~\cite{BC300}
for further details): When the coupling between the 
motion of the mirror and the optical field is zero ($I_o \rightarrow 0$), the  
resonant frequencies of the entire system are given 
by the resonances of the test mass, 
i.e.\ the free-oscillation modes of a test mass ($\Omega =0$),  
plus the resonances of the optical field, i.e.\  
the electromagnetic modes of the entire cavity with fixed mirrors, given by  
$\Omega=\pm \gamma \tan\phi$~\cite{SR}. 
When the light power is increased toward $I_{\rm SQL}$, 
the coupling between the free test mass and 
the optical field drives the four resonant frequencies 
away from their decoupled values. 
By analyzing the four coupled resonant frequencies, we can easily identify 
the ones with the $-$ ($+$) sign in Eq.~(\ref{4.4}) as 
remnants of the resonant frequencies of the free test mass (optical field). 
(For a more thorough discussion of these results see Ref.~\cite{BC300}, 
where we explicitly examine the mirror motion.)

Let us observe that 
Eq.~(\ref{4.3}) can also be obtained as follows.
By expanding the noise spectral density (\ref{3.5}) 
for $\tau \rightarrow 0$, we get:
\beq
\frac{S_h(\Omega)}{h^2_{\rm SQL}(\Omega)}
=  \frac{{\left( -2\,\cos 2\,\beta  + 2\,\cos 2\,\phi  + 
{\cal K}\,\sin 2\,\phi  \right) }^2}{8\,{\cal K}\,
\left [ \cos^2\beta\,(\sin^2 \zeta - \cos^2 \phi) +
\cos^2 \phi\,\cos^2 \zeta \right ] }\, 
\frac{1}{\tau^2}+{ \cal O}(\tau^0)\,.
\label{4.5}
\eeq
The leading term of the expansion goes to zero when 
$2\cos{2\phi}-2\cos{2\beta_{\rm res.}}+{\cal K}\sin{2\phi}=0$,
which is exactly the resonant condition (\ref{4.3}) 
for the closed system derived above.  This means that
for (open) SR interferometers with highly reflecting SR mirrors, the 
dips in the noise curves agree with the resonances of the closed system.
\begin{figure}
\centerline{
\epsfig{file=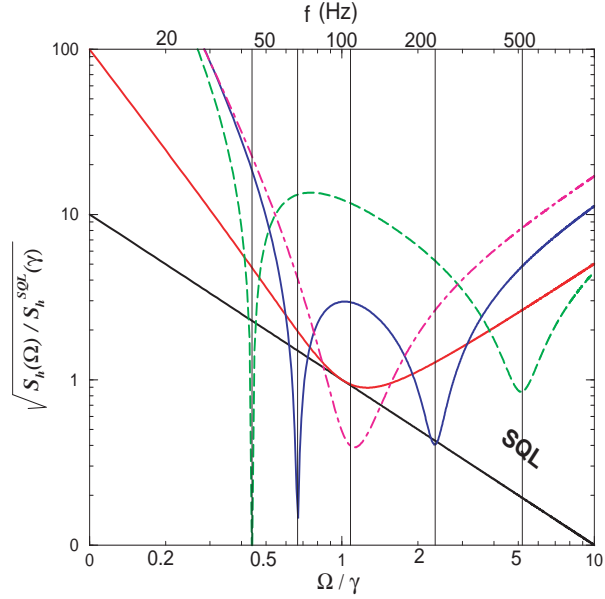,width=0.5\textwidth,height = 0.5\textwidth,angle=-90}}
\vskip 0.2truecm
\caption{\sl Log-log plot of $\sqrt{S_h(\Omega)/S_h^{\rm SQL}(\gamma)}$ 
versus $\Omega/\gamma$ for $I_o=I_{\rm SQL}$, $\rho=0.95$ and 
$\zeta =0$ (i.e.\ 
the second quadrature $b_2^{\cal C}$ is measured). 
The detuning phase $\phi$ takes the values (going from right to left): 
$\pi/2 - 0.19$, $\pi/2 - 0.39$ and $\pi/2 - 0.59$. 
The vertical grid lines have been drawn by using Eq.~(\ref{4.4}) 
and taking the real part of $\Omega_{\rm res.}$.
These lines agree well with the positions of the resonant dips.} 
\label{Fig7}
\end{figure}

In practice, the real part of the resonant frequencies ~(\ref{4.4}) for the closed 
system turns out to be a good approximation  to 
the positions of the valleys in the noise 
spectral density of an (open) SR interferometer with high SR-mirror reflectivity.
To illustrate this fact, in Fig.~\ref{Fig7} we plot 
the noise curves $\sqrt{S_h(\Omega)}$ for the second quadrature $b_2^{\cal C}$ 
with $I_o = I_{\rm SQL}$, $\rho = 0.95$ and 
varying $\phi$. The vertical lines have been drawn 
by solving Eq.~(\ref{4.4}) numerically for $\Omega$  and taking its  
real part, i.e., the real part of the resonant frequencies of the closed systems.
There is indeed very good agreement. This suggests 
that the gain in sensitivity comes from a resonant amplification effect;  
see the discussion at the end of the Sec.~\ref{subsec4.3}.

If the imaginary part of the resonant frequency 
is positive (negative) then, with our convention for the Fourier transform, 
the solution is unstable (stable). The best noise sensitivity curves have detuning phase 
$\phi$ in the range $\arctan[(4I_o/I_{\rm SQL})^{1/3}]\, \laq\, \phi \,\laq\, \pi/2$, which 
for $\rho =1$ correspond to two real resonant frequencies, and 
no instability. However, 
as soon as we allow 
the transmissivity of the SR mirror $\tau$ to be different from zero (as it must 
be in a real interferometer),  
we always find that one of the two resonant frequencies 
has a positive imaginary part \cite{BC300}. A more detailed analysis 
of the dynamics of the system has shown that this is a rather weak 
instability which typically develops on a time scale of $ \laq\, 0.1 \gamma$ 
and can be cured by introducing an appropriate control system \cite{BC300}.

\subsection{Semiclassical interpretation of resonances for small $\cal K$: pure optical resonances}
\label{subsec4.2}

In this section we shall focus on the optical-field resonances 
and shall relate our results to previous semiclassical analyses of 
SR interferometers \cite{SR,M95}.

The test-mass motion affects the optical fields through the term 
${\cal K}= 2\left(I_o/I_{\rm SQL}\right) \gamma^4/\left(\Omega^2 (\Omega^2+\gamma^2)\right)$, where 
the factor $I_o/I_{\rm SQL}$ can be considered a measure of the strength of 
the coupling. The quantity $\cal K$ governs 
both the resonant condition and the relative magnitude of shot noise 
and radiation-pressure noise.
In particular, when ${\cal K}$ is very small, Eq.~(\ref{4.3}) simplifies to   
$\cos{2\phi}-\cos{2\beta_{\rm res.}}=0$, which can be solved 
easily, giving:  
\beq
2(\pm\beta_{\rm res.}+\phi)= 2\pi\,n\,, \quad \quad {\rm i.e.} \quad 
\Omega_{\rm res.}=\pm\gamma\tan\phi\,,
\label{4.6}
\eeq
with $n$ an integer. Equation (\ref{4.6}) can be explained with a simple optics argument:    
The quantity $\pm 2\beta$ is the phase gained by the upper and lower GW
sidebands while in an arm cavity, while $\phi$ is the phase 
gained when traveling one way down the SR cavity. 
Thus $2(\pm\beta+\phi)$ is just the round-trip phase, and Eq.~(\ref{4.6}) is the resonant 
condition for the entire (closed) interferometer. Hence, the presence 
of $\cal K $ in the resonant  condition (\ref{4.3}) provides the 
deviation from a pure optical resonance.
Moreover, $\cal K$ is also an indicator of the different scalings of 
$I_o$ and $m$ in the final expressions for the noises, and therefore it 
governs the relative magnitude of the shot noise and radiation-pressure noise 
-- the smaller the $\cal K$, the more important the 
shot noise compared to radiation pressure noise.
When $\cal K$ is small, a semi-classical argument helps to 
explain the features of our noise curves. 
If we are close to the resonance, then feeding back the signal at that frequency 
increases the peak sensitivity while decreasing the bandwidth.  
Different schemes of such narrow-banding have been proposed, 
e.g., see Drever \cite{D82}.
The scheme discussed here, in which the signal at the dark port is fed
back into the arm cavities,
is called signal recycling (in the narrower sense), 
and was invented  by Meers \cite{SR}. 
If, on the other hand, we are far enough 
from the resonances, sideband signals are not encouraged to go back 
into the interferometer; 
in particular, at $|\beta_{\text{anti-res.}}|\simeq |\beta_{\rm res.}\pm\pi/2|$, 
there is antiresonance, and the signal is encouraged to go out. This is 
what is generally called resonant sideband-extraction and was invented 
by Mizuno \cite{M95}, see Sec.~\ref{subsec3.3}. 
The range in between, $\beta_{\rm res.} < \beta < \beta_{\text{anti-res.}}$,  
is called `detuned' signal recycling and has recently been
demonstrated experimentally on the 30\,m laser interferometer at
Garching, Germany by Freise et al.\ \cite{Freise00} and at Caltech on a
table-top experiment by Mason \cite{M01}.

\begin{figure}
\centerline{
\epsfig{file=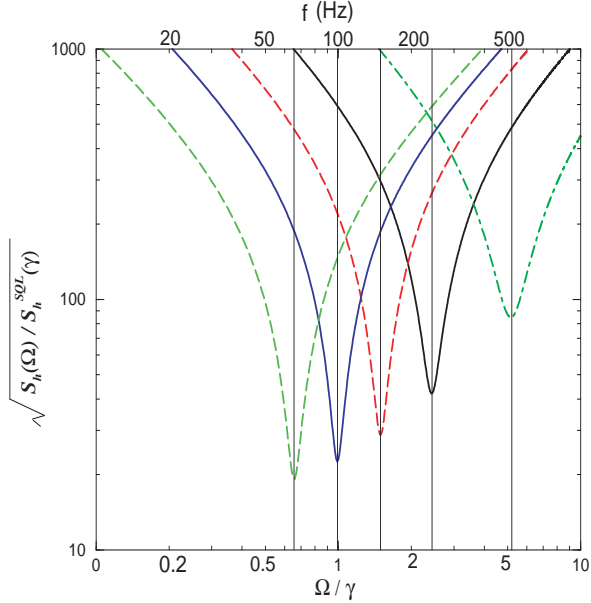,width=0.5\textwidth,height = 0.5\textwidth,angle=-90}}
\vskip 0.2truecm
\caption{\sl Log-log plot of 
$\sqrt{S_h(\Omega)/S_h^{\rm SQL}(\gamma)}$ versus $\Omega/\gamma$ 
for $\zeta=0$ (i.e.\ $b_2^{\cal C}$ is measured) and 
for extremely low light power and high reflectivity: $I_o=10^{-4}I_{\rm SQL}$ and  
$\rho=0.95$. 
$\phi$ takes the values (going from right to left): $\pi/2 - 0.19$, $\pi/2 - 0.39$, 
$\pi/2 - 0.59$, $\pi/2 - 0.79$ and  $\pi/2 - 0.99$. 
A series of resonances appear whose positions agree with the vertical grid lines 
drawn according to ${\Omega_{\rm res.}}/{\gamma}=|\tan\phi|$ [Eq.~(\ref{4.6})].}
\label{Fig6}
\end{figure}

As an example of resonance (not anti-resonance), we plot 
in Fig.~\ref{Fig6} the spectral density 
$S_h(\Omega)$ when the second quadrature $b_2^{\cal C}$ is measured,  
for very low light power $I_o = 10^{-4}\,I_{\rm SQL}$ 
and high reflectivity $\rho = 0.95$, and for various values 
of the detuning phase $\phi$. 
The vertical grid lines in Fig.~\ref{Fig6} are drawn according to 
Eq.~(\ref{4.6}) and indeed, there is excellent agreement. 

It is interesting to note that although for LIGO-II $I_o = I_{\rm SQL}$, 
there is still a frequency band where $\cal K$ is relatively small. 
This is due to the fact that $\cal K$ drops very fast as 
$\Omega$ increases. In that frequency band the semiclassical 
formalism gives a correct result for the optical resonances \cite{KS}. 
However, since the semiclassical approach 
does not take into account the motion of the arm-cavity end mirrors, 
it can only describe one resonance (and not two) in the entire spectrum.

\subsection{Quantum mechanical discussion of the general case: two resonances and $\rho \neq 1$}
\label{subsec4.3}
The correspondence between the optical-mechanical resonances and the minima 
of the noise curves suggests that the gain in sensitivity comes 
from a resonant amplification of the input signal, i.e.\ of the 
gravitational force acting on the mirrors, 
as already observed for optical bar GW detectors by Braginsky's group 
\cite{OB}. Let us discuss this point more deeply. 

The quantum part of the input--output relation ~(\ref{2.20}) 
(with $|\Phi| \ll 1$ as we have assumed throughout this paper) 
reads 
\beq
\label{4.7}
b_i^{\rm quant} = \frac{e^{2 i \beta}\,C_{i j}}{M}\,a_j \,, \quad \quad 
i,j = 1,2\,.
\eeq
We find it convenient to renormalize the quantum transfer matrix:
\beq
\label{4.8}
{\cal M}_{i j} \equiv  \frac{C_{i j}}{|M|}\,, \quad \quad 
i,j = 1,2
\eeq
so $\det{\cal M}_{i j} = 1$. Note that this ${\cal M}_{i j}$ 
is normalized with respect to unit quantum noise.
Because the $C_{i j}$ are real, the matrix ${\cal M}$ 
depends on three real parameters and we can always decompose it  
into two rotations $R(\theta)$, $R(\varphi)$ and a squeeze $S(r)$ 
(see for details Ref.~\cite{CS285}), 
e.g.,  ${\cal M} = R(\theta)\,R(\varphi)\,{S}(r)\,R(-\varphi)$, with 
\beq
R(\theta) = \left(\matrix{ \cos \theta & -\sin \theta \cr \sin \theta  & \cos \theta}\right) \,,
\quad \quad 
{S}(r) = \left(\matrix{e^r
& 0  \cr  0 & e^{-r}}\right) \,,
\label{4.9}
\eeq
where the factor $e^r$ describes the stretching ($r>0$)
or squeezing ($r <0$) of the quantum fluctuations 
in the quadrature $b_i$  [see Eqs. (\ref{4.7}), (\ref{4.8})].  
Note that classical optical fields always have a zero squeeze factor. 

To express the squeeze parameter $r$ in terms of the physical parameters 
describing the SR interferometer, we simply take the trace of the matrix 
${\cal M}\,{\cal M}^{\dagger}$, obtaining
\beq
\label{4.10}
e^{2 r} + e^{-2r} = 2 + \frac{\tau^4\,{\cal K}^2}{|M|^2}\,.
\eeq
Hence, in a SR interferometer the squeezing (generally called 
ponderomotive squeezing) is induced by the 
back-action force acting on the mirror through the effective 
coupling ${\cal K}$. 
In particular, for small ${\cal K}$, we have $e^{2r}+e^{-2r} \approx 2$
and the squeeze factor $r$ goes to $0$, which means the output
field is classical. For our discussion below the 
specific expressions of $\theta$ and $\varphi$ in terms of the physical 
parameters are unimportant. 

{}From the previous discussions and the results derived in \cite{BC300} 
we have learned that the zeros of $M(\Omega)$ are the resonant 
frequencies of the optical-mechanical system 
and the valleys of the noise spectral densities are their real parts. 
It is straightforward to show that for $\Omega$ equal 
to the real part of the resonances $|M| \propto \tau^2$. 
Hence, on resonance, for typical values of the physical quantities $I_o$, $\rho$ and 
$\phi$, the RHS of Eq.~(\ref{4.10}) goes to a constant when 
$\tau \rightarrow 0$. This means that the squeeze factor 
$r$ does not grow much around the resonances. 
On the other hand, the absolute value of the output signal 
strength [the term involving $h$ in Eq.~(\ref{2.20})], is given by 
\beq
\frac{\sqrt{2 {\cal K}}\,\tau\,|D_i|}{h_{\rm SQL}\,|M|}\,h\,, 
\quad \quad i = 1,2\,,
\label{4.11}
\eeq
and because on resonance $1/|M| \! \sim \! 1/\tau^2$, when $\tau \rightarrow 0$ 
the classical signal is resonantly amplified and the amplification 
becomes stronger and stronger as $\tau \rightarrow 0$ (closed system). 
 
This means that, by contrast with QND techniques based on 
static correlations between shot noise 
and radiation-pressure noise \cite{KLMTV00}, \cite{HFD}, 
in SR interferometers the ponderomotive squeezing 
\emph{does not} seem to be the major factor that enables the interferometer 
to beat the SQL. Indeed, whereas the amplitude of the classical 
output signal is amplified near the resonances, 
the nonclassical behavior of the output light is not resonantly amplified.
Therefore, the beating of the SQL in SR interferometers 
comes from a resonant amplification of the input signal: 
the whole system acts like an optical spring,\footnote{~In this sense
we could refer to a signal recycled interferometer  
as a SPRING detector, which could also stand for Signal 
Power Recycling Interferometer Gravitational detector !} 
as we have described more thoroughly in \cite{BC300}, and 
it was also derived for optical bar GW detectors by Braginsky's group 
\cite{OB}.

\section{Inclusion of losses in signal-recycling interferometers}
\label{sec5}
In this section we shall compute how optical losses affect the noise in 
a SR inteferometer using the lossy input--output relations for 
a conventional interferometer \cite{KLMTV00} and doing a similar treatment 
of losses in the SR cavity. We shall continue to use 
our extension of the KLMTV's formalism as developed in Sec.~\ref{sec2}. 
In \cite{BC300} we show that when losses are included a suitable control system 
can be implemented to circumvent the instabilities.

\begin{figure}
\begin{center}
\vspace{-0.8cm} 
\epsfig{file=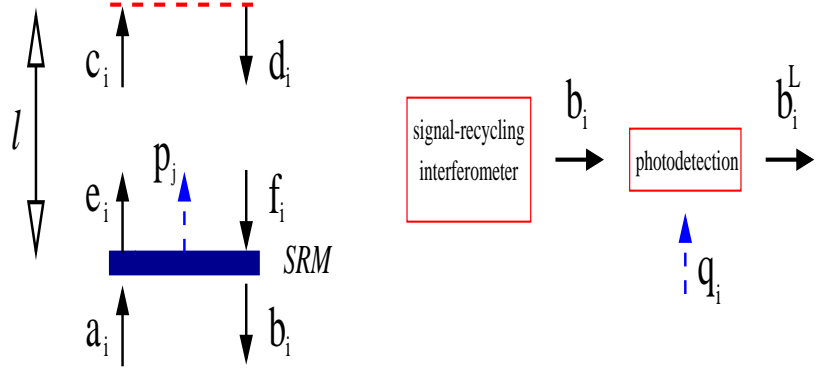,width=0.3\textwidth,height = 0.65\textwidth,angle=-90}
\vskip 0.2truecm
\caption{\sl Sketchy view of the lossy signal-recycling interferometer. 
Optical losses in the signal-recycling cavity (on the left) 
are described by the noise quadratures $p_i$, while 
losses due to the photodetection process (on the right) are included 
through the noise quadratures $q_i$.} 
\label{Fig8}
\end{center}
\end{figure}

KLMTV~\cite{KLMTV00} described the noise that enters the arm cavities 
of a conventional interferometer at the loss points on the mirrors 
in terms of a noise operator, whose state is the vacuum, 
with quadratures $n_1$ and $n_2$.
The resulting lossy input--output relations read \cite{KLMTV00}
\bea
\label{5.1}
d_1 &=& c_1 \,e^{2 i \beta}\, \left ( 1 - \frac{{\cal E}}{2} \right )+ 
\sqrt{{\cal E}}\,e^{i \beta}\,n_1\,, \\
d_2 &=& c_2 \,e^{2 i \beta}\, \left ( 1 - \frac{{\cal E}}{2} \right )+ 
\sqrt{{\cal E}}\,e^{i \beta}\,n_2 + 
\sqrt{2 {\cal K}}\,\frac{h}{h_{\rm SQL}}\,e^{i \beta}\,
\left [ 1 - \frac{\epsilon}{4}\,(3 + e^{2 i \beta}) \right ] \nonumber \\
&& - {\cal K}\, e^{2 i \beta}\,\left \{ c_1\,\left [ 
1 - \frac{\epsilon}{2}\,(3 +  e^{2 i \beta}) \right ] + 
\sqrt{\frac{\epsilon}{2}}\,n_1 \right \}\,,
\label{5.2}
\eea
where $\epsilon = 2 {\cal L}/T$ and ${\cal L}$ is the loss coefficient 
per round trip in the arm-cavity. For LIGO-II $T$ and ${\cal L}$ 
are expected to be $T = 0.033$ and ${\cal L} \! \sim \!  200 \times 10^{-6}$, 
so $\epsilon \! \sim \! 0.01$. The quantity ${\cal E}$ which 
appears in Eqs.~(\ref{5.1}) and  (\ref{5.2}) is frequency dependent and 
is given by
\beq
{\cal E} = \frac{2 \epsilon}{1 + (\Omega/\gamma)^2}\,.
\label{5.3}
\eeq
In the present analysis, as in Ref.~\cite{KLMTV00}, we do not take into account losses coming 
from the beam splitter. We expect their effect to be 
small compared to the losses introduced by the 
SR cavity and the photodetection process. 
Fig.~\ref{Fig8} sketches the way we have incorporated losses. 
We describe the loss inside the SR 
cavity by the fraction of photons lost  at each bounce
of the interior field off the SR mirror, $\lambda_{\rm SR}$, 
and we introduce associated noise quantum operators $p_i$ ($i=1,2$) 
into the inward-propagating field operator at the SR mirror 
(see left panel of Fig.~\ref{Fig8}). Equations (\ref{2.18}) then become
\beq
e_1 = \sqrt{1 -\lambda_{\rm SR}}\,(\tau\,a_1+\rho\, f_1)  + 
\sqrt{\lambda_{\rm SR}}\,p_1 \,, \quad \quad 
e_2 = \sqrt{1 -\lambda_{\rm SR}}\,(\tau\,a_2+\rho\, f_2) + 
\sqrt{\lambda_{\rm SR}}\,p_2 \,,
\label{5.4}
\eeq
and the noise operators $p_i$ satisfy 
the commutation relations (\ref{com}). We also assume that the state of $p_i$ 
is the vacuum. We include the losses of the photodetection process 
in an effective way, by modifying the 
output field operators and introducing another noise field $q_i$ with $i = 1,2$ (see 
right panel of Fig.~\ref{Fig8}):
\beq
b^{\rm L}_1= \sqrt{1 -\lambda_{\rm PD}}\,(\tau\, f_1-\rho\,a_1)  + 
\sqrt{\lambda_{\rm PD}}\,q_1\,, \quad \quad 
b^{\rm L}_2 = \sqrt{1 -\lambda_{\rm PD}}\,(\tau\, f_2-\rho\,a_2)  + 
\sqrt{\lambda_{\rm PD}}\,q_2\,.
\label{5.5}
\eeq
Here, $\lambda_{\rm PD}$ is the photodetector loss. 
The noise quadrature fields $q_i$ describe additional shot noise 
due to photodetection and are assumed to satisfy Eq.~(\ref{com}) and 
to be in the vacuum state. 
Following the procedure described in Sec.~\ref{sec2},   
we derive from Eqs.~(\ref{5.1}), (\ref{5.2}), (\ref{5.4}) 
and (\ref{5.5}) the following input--output relations 
for the lossy SR interferometer (for simplicity we set $\Phi = 0$):
\bea
\left (\matrix{ b^{\rm L}_1 \cr b^{\rm L}_2}\right)&=&
\frac{1}{M^{\rm L}}\left[e^{2i\beta}\,\left(\matrix{ C^{\rm L}_{11} & C^{\rm L}_{12}\cr 
C^{\rm L}_{21} & C^{\rm L}_{22}}\right)
\left(\matrix{ a_1\cr a_2}\right)+
\sqrt{2 {\cal K}}\,\tau\,e^{i\beta}\,\left(\matrix{ D^{\rm L}_1\cr D^{\rm L}_2}\right)
\frac{h}{h_{\rm SQL}} + e^{2i\beta}\, \left (\matrix{ P_{11} & P_{12}\cr P_{21} & P_{22}}\right)
\left(\matrix{ p_1 \cr p_2}\right) \right . \nonumber \\
&& \left . + e^{2i\beta}\,
\left(\matrix{ Q_{11} & Q_{12}\cr Q_{21} & Q_{22}}\right)
\left(\matrix{ q_1 \cr q_2}\right) + 
e^{2i\beta}\,\left(\matrix{ N_{11} & N_{12}\cr N_{21} & N_{22}}\right)
\left(\matrix{ n_1\cr n_2}\right) \right ]\,,
\label{5.6}
\eea
where, to ease the notation, we have defined 
\bea
M^{\rm L} 
&=& 1+\rho^2\, e^{4i\beta}-
2\rho\,\left (\cos{2\phi}+\frac{{\cal K}}{2}\,
\sin{2\phi} \right )\,e^{2i\beta} 
+ \lambda_{\rm SR}\,\rho\,\left ( -\rho\,e^{2i\beta}  + \cos{2\phi}+\frac{{\cal K}}{2}\,
\sin{2\phi}\right )\,e^{2i\beta} \nonumber \\
&& + \epsilon\,\rho\,\left [ 2 \cos^2 \beta\,(-\rho\,e^{2i\beta} + \cos2 \phi) + 
\frac{{\cal K}}{2}\,(3 + e^{2 i\beta})\,\sin 2 \phi \right ]\,e^{2i\beta}\,.
\label{5.7}
\eea
Note that $M^{\rm L}$, like $M$ in Eq.~(\ref{2.21}), 
has zeros in the lower- and upper-half complex $\Omega$ plane. Hence, 
the lossy SR interferometer, like the lossless one, 
also suffers from instabilities. Nevertheless,  
we have shown in \cite{BC300} that an appropriate control system can cure them, 
as in the lossless case. 
In the following equations we give the various quantities which 
appear in Eq.~(\ref{5.6}) accurate to linear order in 
$\epsilon$ and $\lambda_{\rm SR}$ but to all orders in 
$\lambda_{\rm PD}$. (We expect $\lambda_{\rm SR} \! \sim \! 0.02$ and 
$\lambda_{\rm PD} \! \sim \!0.1$~\cite{KS}.)
The various quantities read
\bea
C^{\rm L}_{11}&=&C^{\rm L}_{22}=\sqrt{1 - \lambda_{\rm PD}}\,\left \{ (1+\rho^2)\,
\left (\cos{2\phi}+\frac{{\cal K}}{2}\,\sin{2\phi} \right ) -2\rho\,\cos{2\beta}\right . 
\nonumber \\
&& \left . -\frac{1}{4} \epsilon \,\left [ -2(1 + e^{2i\beta})^2\,\rho + 4(1 + \rho^2)\cos^2\beta\,
\cos 2 \phi + (3 + e^{2i\beta})\,{\cal K}\,(1 + \rho^2)\,\sin 2 \phi \right ]\right . \nonumber \\ 
&&+ \left . \lambda_{\rm SR}\,\left [ e^{2i\beta}\rho - \frac{1}{2}\,(1 + \rho^2)\,
\left (\cos{2\phi}+\frac{{\cal K}}{2}\,\sin{2\phi} \right ) \right ]\right \}\,,  \nonumber\\
C^{\rm L}_{12}&=&\sqrt{1 - \lambda_{\rm PD}}\,\tau^2\,\left \{
- (\sin{2\phi}+{\cal K}\,\sin^2{\phi}) + \frac{1}{2}\epsilon\,\sin \phi\,
\left [(3 + e^{2i\beta})\,{\cal K}\,\sin \phi + 4 \cos^2 \beta\,\cos \phi 
\right ] \right . \nonumber \\
&& \left . + \frac{1}{2}\lambda_{\rm SR}\,(\sin 2 \phi + {\cal K}\,\sin^2 \phi ) \right \}\,,  \nonumber \\
C^{\rm L}_{21}&=&\sqrt{1 - \lambda_{\rm PD}}\,\tau^2\,
\left \{ (\sin{2\phi}-{\cal K}\,\cos^2{\phi}) + 
\frac{1}{2}\epsilon\,\cos \phi\,\left [(3 + e^{2i\beta})\,{\cal K}\,\cos \phi - 4
\cos^2 \beta \, \sin \phi \right ] \right . \nonumber \\
&& \left . + \frac{1}{2}\lambda_{\rm SR}\,(-\sin 2 \phi + {\cal K}\,\cos^2 \phi )  \right \}\,;
\label{5.8}
\eea
\bea
D^{\rm L}_1&=& \sqrt{1 - \lambda_{\rm PD}}\,\left \{ -(1+\rho\, e^{2i \beta})\,\sin{\phi}
+ \frac{1}{4}\epsilon\,\left [3 + \rho + 2\rho\, e^{4i\beta} \right . \right .
\nonumber \\
&& \left. \left . + 
e^{2i\beta}\,(1 + 5\rho)\right ]\,\sin \phi +  
\frac{1}{2}\lambda_{\rm SR}\,e^{2i\beta}\,\rho\sin \phi \right \}\,, \nonumber \\
D_2^{\rm L}&=& \sqrt{1 - \lambda_{\rm PD}}\,\left \{
- (-1+\rho\, e^{2i\beta})\,\cos{\phi} + 
\frac{1}{4}\epsilon\,\left [-3 + \rho + 2\rho\,e^{4i\beta} \right. \right. \nonumber \\
&& \left. \left. + 
e^{2i\beta}\,(-1 + 5\rho)\right ]\,\cos \phi +  
\frac{1}{2}\lambda_{\rm SR}\,e^{2i\beta}\,\rho\cos \phi \right \}\,;
\label{5.9}
\eea
\bea
P_{11} &=& P_{22} = \frac{1}{2} \sqrt{1 - \lambda_{\rm PD}}\,
\sqrt{\lambda_{\rm SR}}\,\tau\,(-2\rho\, e^{2i\beta} + 2 \cos 2\phi + 
{\cal K}\,\sin 2 \phi)\,, \nonumber \\
P_{12}&=& - \sqrt{1 - \lambda_{\rm PD}}\,
\sqrt{\lambda_{\rm SR}}\,\tau\,\sin \phi\,(2\cos \phi + {\cal K}\,\sin \phi)\,, 
\nonumber \\
P_{21}&=& \sqrt{1 - \lambda_{\rm PD}}\,
\sqrt{\lambda_{\rm SR}}\,\tau\,\cos \phi\,(2\sin \phi - {\cal K}\,\cos \phi)\,;
\label{5.10}
\eea
\bea
Q_{11} &=& Q_{22} =  \sqrt{\lambda_{\rm PD}}\, \left \{ 
e^{-2i\beta} + \rho^2\,e^{2i\beta}-\rho\,(2 \cos 2 \phi + {\cal K}\,\sin 2 \phi)
+ \frac{1}{2}\epsilon\,\rho\,\left [ 
e^{-2i\beta}\,\cos 2 \phi \right . 
\right. \nonumber \\
&& \left. \left.  + e^{2i\beta}\,\left (-2\rho 
-2\rho\,\cos 2 \beta + 
\cos 2 \phi + {\cal K}\,\sin 2\phi \right ) + 2 \cos 2 \phi + 3 {\cal K}\,\sin 2 \phi 
\right ] \right . \nonumber \\
&& \left. 
-  \frac{1}{2}\lambda_{\rm SR}\,\rho\,\left [ 2\rho\,e^{2i\beta} -2 \cos 2 \phi 
- {\cal K}\,\sin 2 \phi \right ] \right \}\,, \nonumber\\
Q_{12} &=& 0 = Q_{21}\,;
\label{5.11}
\eea
\bea
N_{11} &=& \sqrt{1 - \lambda_{\rm PD}}\,\sqrt{\frac{\epsilon}{2}}\,
\tau\,\left \{ {\cal K}\,(1 + \rho\,e^{2i\beta})\,\sin \phi +
2 \cos \beta \,\left [e^{-i\beta} \cos \phi  - \rho e^{i\beta}
(\cos \phi + {\cal K}\,\sin \phi ) \right ] \right \}\,, \nonumber \\
N_{22} &=& - \sqrt{1 - \lambda_{\rm PD}}\,\sqrt{2\epsilon}\,
\tau\,(-e^{-i\beta} + \rho\,e^{i\beta})\,\cos \beta\,\cos \phi\,, \nonumber \\
N_{12} &=&  - \sqrt{1 - \lambda_{\rm PD}}\,\sqrt{2\epsilon}\,
\tau\,(e^{-i\beta} + \rho\,e^{i\beta})\,\cos \beta\,\sin \phi\,, \nonumber \\
N_{21} &=& \sqrt{1 - \lambda_{\rm PD}}\,\sqrt{\frac{\epsilon}{2}}\,
\tau\,\left \{ -{\cal K}\,(1 + \rho)\,\cos \phi + 
2 \cos \beta \,\left (e^{-i\beta} + \rho\, e^{i\beta} \right )\,
\cos \beta \, \sin \phi  \right \}\,.
\label{5.12}
\eea
Similarly to Sec.~\ref{subsec3.1}, we follow KLMTV's method \cite{KLMTV00} 
to derive the noise spectral density of a lossy SR interferometer [see Eq.~(\ref{3.5})]:
\bea
\label{lossy}
S_h^{\zeta}= \frac{h_{\rm SQL}^2}{2{\cal K}\,\tau^2\,\left|D^L_1\,\sin\zeta+D^L_2\,\cos\zeta\right|^2}\,
&&\left [\left |C^L_{11}\,\sin\zeta+C^L_{21}\,\cos\zeta\right |^2+
\left |C^L_{12}\,\sin\zeta+C^L_{22}\,\cos\zeta\right |^2 +\right. \nonumber \\
&& \left |P_{11}\,\sin\zeta+P_{21}\,\cos\zeta\right |^2+
\left |P_{12}\,\sin\zeta+P_{22}\,\cos\zeta\right |^2+   \nonumber \\
&&\left |Q_{11}\,\sin\zeta+Q_{21}\,\cos\zeta\right |^2+
\left |Q_{12}\,\sin\zeta+Q_{22}\,\cos\zeta\right |^2+   \nonumber \\
&&\left. \left |N_{11}\,\sin\zeta+N_{21}\,\cos\zeta\right |^2+
\left |N_{12}\,\sin\zeta+N_{22}\,\cos\zeta\right |^2 
\right ]\,.
\eea

\begin{figure}
\centerline{\epsfig{file=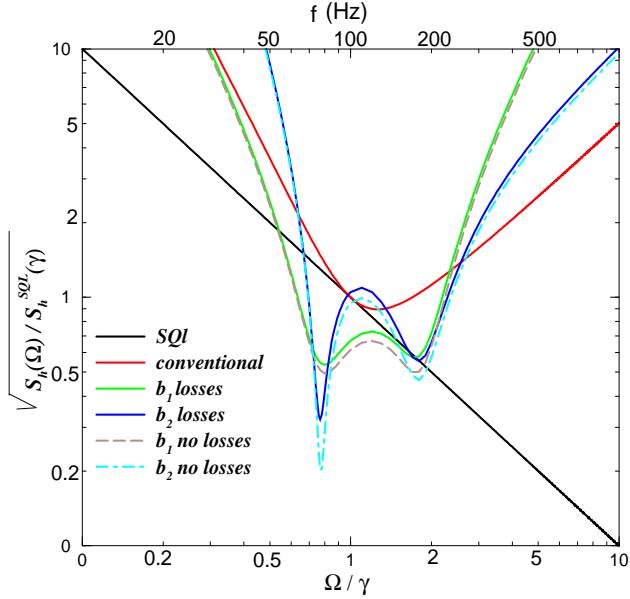,width=0.5\textwidth,height = 0.5\textwidth,angle=-90}}
\vskip 0.1truecm
\caption{Log-log plot 
of $\sqrt{S_h(\Omega)/S_h^{\rm SQL}(\gamma)}$ versus $\Omega/\gamma$ for the 
two quadratures $b_1$ ($\zeta = \pi/2$)and $b_2$ 
($\zeta = 0$), including and not including losses, 
with $\rho = 0.9$, $\phi = \pi/2 -0.47$, $I_o = I_{\rm SQL}$, 
$\epsilon = 0.01$, $\lambda_{\rm SR} = 0.02$ and $\lambda_{\rm PD} = 0.1$. 
The noise curve for a conventional 
interferometer and the SQL are shown as well.}
\label{Fig9}
\end{figure}

Exploring numerically this equation, we find that for the loss levels expected in 
LIGO-II ($\epsilon \! \sim \! 0.01, \lambda_{\rm PD} \! \sim \! 0.1, \lambda_{\rm SR} 
\! \sim \! 0.02$ \cite{GSSW99}), 
the optical losses have only a modest influence on the noise curves 
of a lossless SR interferometer. For example, 
in Fig.~\ref{Fig9} we compare the lossless noise spectral densities 
with the lossy ones for the two quadratures $b_1$ and $b_2$. 
The main effect of the loss is to  smooth out the deep resonant valleys. 
More specifically, for (i) the physical parameters used in Fig.~\ref{Fig2}, 
(ii) a net fractional photon loss of 1\% in the arm cavities ($\epsilon = 0.01$) 
and 2\% in each round trip in the SR cavity ($\lambda_{\rm SR} = 0.02$), 
and (iii) a photodetector efficiency of 90\% ($\lambda_{\rm PD} =0.1$),
we find that the losses produce a fractional loss in signal-to-noise ratio for 
inspiraling binaries [see Eqs.~(\ref{3.8}), (\ref{3.9})] 
of $8\%$ and $21\%$, for the first and second quadratures, respectively.

The reason why we get a modest effect from optical losses as compared to schemes
using squeezing or FD homodyne detections\footnote{~Note that in KLMTV \cite{KLMTV00} 
they assumed a loss factor for end-mirrors which is $10\%$ of our value, 
and they also did not take into account losses coming from the 
photodetection.} rests on the 
fact that our gain in sensitivity mostly comes from resonant amplification, 
which is much less susceptible to losses than quantum correlations. This
general consideration has long been understood by Braginsky, Khalili and colleagues
and underlies their motivation for the GW ``optical bar'' detectors \cite{OB}.

\section{Conclusions}
\label{sec6}
In this paper we have extended the quantum formalism recently 
developed \cite{KLMTV00} for conventional interferometers 
(LIGO-I/TAMA/Virgo), to SR interferometer such as LIGO-II. 
The introduction of the SR cavity has been planned as an important 
tool to reshape the noise curves, making the interferometer 
work either in broadband or in narrowband configurations. 
This flexibility is expected to improve the observation of 
specific GW sources \cite{Sour}. 
Quite remarkably, our quantum mechanical analysis 
has revealed other significant features of the SR cavity. 

First, the SR mirror produces dynamical correlations 
between quantum shot noise and radiation-pressure-fluctuation 
noise which break the light's ability to enforce the SQL of a free mass, 
allowing the noise curves to go below the SQL by modest amounts: 
roughly a factor two over a bandwidth $\Delta f \!\sim\! f$. 
Before our work, researchers were unaware of  
the shot-noise / radiation-pressure correlations 
and thus omitted them in their semiclassical analysis 
of the straw-man design of LIGO-II \cite{GSSW99}.
The goal of beating the SQL in LIGO-II can be achieved 
only {\it if} all sources of thermal noise can also be
pushed below the SQL and indeed much R\&D will go into 
trying to push them downward. It turns out that even with current 
estimates of the LIGO-II thermal noise \cite{BGV00}, which are a little 
above the SQL, the net noise (thermal plus optical) is significantly 
affected by the shot-noise/ radiation-pressure correlations. 
Indeed, the  correlations lift the noise at low frequencies   
$ 10 \, {\rm Hz} \,\laq\, \Omega/ 2\pi \,\laq\, 50$ Hz,
as compared to the semiclassical estimations, even though in this 
frequency range the optical noise 
may already be very much larger than the SQL. This is due to 
the inaccuracy of the semiclassical method in estimating the effect
of the radiation-pressure force, which is important in this region.
In the middle frequency range, i.e.\ near 100\,Hz, the SQL-beating
effect cannot lower the total noise much because of the thermal contribution.  
The effect of the correlations in the implementation of LIGO-II 
will be clarified and sharpened once the readout scheme 
has be specified \cite{BC400}. 

Second, we have learned that the \emph{dynamical} correlations 
arise naturally from the nontrivial coupling between the antisymmetric 
mode of motion  of the four arm-cavity mirrors and the signal recycled 
optical fields. This dynamical coupling invalidates 
the naive picture, according to which the arm cavity mirrors are subject only to random 
quantum-vacuum fluctuations. The SR interferometer
responds to a GW signal as an optical spring \cite{BC300}, 
and this oscillatory response gives the possibility for resonant 
amplification of the GW signal. 
The optical-mechanical system is characterized by two 
resonances and one of them is always unstable, so 
a control system must be introduced to stabilize it \cite{BC300}.
In the limit of a highly reflecting SR mirror 
we have worked out analytically a very simple equation which 
locates the positions of the resonant frequencies. 
Whereas the amplitude of the classical 
output signal is amplified near the resonances, 
the quantum noise is not particularly affected by them.
All this suggests that the beating of the SQL 
in SR interferometers comes primarily from the resonant 
amplification of the input GW signal, as also occurs in 
``optical bar'' GW detectors \cite{OB}.

The inclusion of losses does not greatly affect 
the SR interferometer. This is due to the fact that 
the improvement in the noise curves rests primarily 
on a resonant amplification and only modestly on ponderomotive 
squeezing. It is worthwhile to point out that the  
SR interferometers bears strong similarity to the ``optical bar'' 
detectors proposed by Braginsky, Khalili and colleagues \cite{OB}. Both
of them can be viewed as oscillators with two different eigenfrequencies.
However, because in SR interferometers the light plays 
the double role of providing the restoring force
and being a probe to monitor the mirror displacements,  
we are forced to introduce in SR interferometers much higher
laser power, to circulate
in the arm cavities ($ \sim \! 1\,{\rm MWatt}$), 
than in the ``optical bar'' scheme. Nevertheless, like the 
``optical bar'' scheme, the SR interferometer
is still less susceptible to optical losses than many other schemes 
designed to beat the SQL. 

It is now important to identify the best SR configuration, 
i.e.\ the choice of the physical parameters   
(light power $I_o$, SR detuning $\phi$, 
reflectivity of SR mirror $\rho$, quadrature phase $\zeta$, 
and the read-out scheme: homodyne or
modulation/demodulation) that optimizes the signal-to-noise 
ratio for inspiraling binaries, for low-mass X-ray binaries, and for  
other astrophysical GW sources. 
We shall discuss this issue in a forthcoming paper 
\cite{BC400}.

Finally, our analysis has shown that dynamical correlations, i.e.\ correlations 
that are intrinsic to the dynamics of the test mass-optical field system
(i.e.\ they are not due to specific read-out schemes, 
as in the case of homodyne detection on a conventional interferometer), 
are present 
when the carrier frequency $\omega_o$ is detuned from 
resonance ($\phi \neq 0$) or antiresonance 
($\phi \neq \pi/2$) in the SR cavity. This suggests a speculation that it could be 
worthwhile to investigate a LIGO-II configuration (see Table \ref{Tab1}) 
without a signal recycling mirror, in which the correlations are 
produced by detuning the arm cavities. However, this case 
will require a very careful analysis of the radiation-pressure force 
acting on the arm-cavity mirrors \cite{PDHV00}, \cite{HR00}.

\acknowledgments 
We wish to thank Yu.~Levin, J.~Mason, N.~Mavalvala and 
K.~Strain for very helpful and stimulating discussions and comments.
It is also a pleasure to thank V.~Braginsky for pointing 
out the importance of optical-mechanical oscillations 
in GW detectors and F.Ya.~Khalili for very 
useful interactions on the optical-mechanical rigidity 
present in LIGO-II. We thank A.~R\"udiger for
his warm encouragements and his very careful reading of the manuscript.
Finally, we are deeply indebted to K.S.~Thorne for 
his constant support and for pointing out 
numerous useful comments and suggestions.

This research was supported by NSF grant PHY-9900776 and 
for AB also by Caltech's Richard Chace Tolman Foundation.

\newpage
\appendix
\section{Remark on commutation relations among 
quadrature fields in Caves-Schumaker two-photon formalism}
\label{appA}
As originally pointed out by Braginsky's group \cite{BK92}  
and discussed by BGKMTV \cite{BGKMTV00}, the output variables  
of the GW interferometer should commute with themselves 
at different times, to guarantee that no other quantum noise is 
necessarily introduced into the measurement 
result once further manipulations
are performed on the output.
Indicating generically by ${\cal O}(t)$ the output 
quantity, the following conditions should be satisfied, 
\beq
[{\cal O}(t),{\cal O}(t')]=0 \quad \quad \forall \, t,t'\,,     
\quad \quad \Leftrightarrow \quad \quad 
[{\cal O}(\Omega),{\cal O}^\dagger(\Omega')]=0 
\quad \quad \forall \,\Omega,\Omega'\,.     
\label{a1}
\eeq
If we assume that the system's output is one quadrature 
of the quantized electromagnetic field (EM) [see Eq.~(\ref{2.10})], 
with the GW signal encoded at side-band frequency 
$\Omega$ around the carrier frequency 
$\omega_o$, then the presence of terms proportional to $\Omega/\omega_o$ 
in Eq.~(\ref{2.4}) prevents the output quadratures 
from commuting with themselves at different times. 
However, Braginsky et al.~\cite{BGKMTV00} anticipated that, 
in the case of LIGO-I/TAMA/Virgo, the quadrature fields 
at the dark port should anyway 
satisfy very accurately the Fourier-domain condition 
given by Eq.~(\ref{a1}), because the side-band frequency 
$\Omega$ ($1\,{\rm Hz}\leq \Omega/2\pi \leq 10^3\,{\rm Hz}$) 
is much smaller than the carrier frequency 
$\omega_o$ ($\omega_o \! \sim \! 10^{15}\,\rm s^{-1}$). 
In this Appendix we investigate this approximation in much more detail, 
estimating the amount of extra noise which will be present in the 
final noise spectral density as a result of 
condition (\ref{a1}) being violated. Henceforth, 
for simplicity we restrict our analysis to conventional 
interferometers.

If the readout scheme is implemented by photodetection,
then only a small frequency band around $\omega_o$ contains 
the final output signal. Hence, it is 
physically justified to introduce a cutoff $\Lambda$ 
in the frequency domain which automatically discards all 
the Fourier components of the EM field outside the range   
$\left[\omega_o-\Lambda,\,\omega_o+\Lambda\right]$ with $0\le\Lambda\le\omega_o$.
As a consequence, Eq.~(\ref{2.6}) for the EM field can be rewritten as 
[see also Eqs. (4.22) of Ref.~\cite{CS185}]
\bea
{E}^{\Lambda}(t)&\equiv&
\int_{\omega_o-\Lambda}^{\omega_o+\Lambda}\,
\sqrt{\frac{2\pi\hbar\omega}{{\cal A} c}}\,
a_{\omega}\,e^{-i \omega t}\, \frac{d\omega}{2\pi}\,+\,{\rm h.c.}  \nonumber \\ 
&=&\sqrt{\frac{2\pi\hbar\omega_o}{{\cal A} c}}\,e^{-i \omega_o t}\,
\int_0^{\Lambda}\frac{d\Omega}{2\pi}
\left( {a}_+ \,e^{-i \Omega t}+{a}_-\, e^{+i \Omega t}\right)
\,+\,{\rm h.c.}	\nonumber\\
&=&\sqrt{\frac{4\pi\hbar\omega_o}{{\cal A} c}}
\left[\cos(\omega_o t)\,{{\cal O}}^{\Lambda}_1(t)+
\sin(\omega_o t)\,{{\cal O}}^{\Lambda}_2(t)\right]\,,
\label{a2}
\eea
where $a_+(\Omega)$ and $a_-(\Omega)$, with $\Omega < \Lambda$, are 
defined by Eq.~(\ref{2.3}) and the 
{\it rescaled} quadrature fields ${{\cal O}}^{\Lambda}_i(t)$ are 
\beq
\label{a3}
{{\cal O}}^{\Lambda}_i(t)\equiv\
\int_0^{\Lambda}
\frac{d\Omega}{2\pi}
\left[{a}_i\, e^{-i \Omega t}+{a}^\dagger_i\, e^{i \Omega t}\right] 
\quad \quad i=1,\,2\,,
\eeq
with the quadrature operators given by Eq.~(\ref{2.7}). 
Evaluating the commutation relations among the quadrature operators we find 
[see also Eqs. (4.31) of Ref.~\cite{CS185}]:
\bea
\label{a4}
\left[{a}_1,{a}_{1'}\right]=\left[{a}_2,{a}_{2'}\right]
&=&0\,, \\
\label{a5}
\left[{a}_1,{a}^{\dagger}_{1'}\right]=\left[{a}_2,{a}^{\dagger}_{2'}\right]
&=&2\pi\delta(\Omega-\Omega')\left(\frac{\Omega}{\omega_o}\right)\,, \\
\label{a6}
\left[{a}_1,{a}^{\dagger}_{2'}\right]=-\left[{a}_2,{a}^{\dagger}_{1'}\right]
&=&2\pi i \delta(\Omega-\Omega') \,.
\eea
Note that Eq.~(\ref{a5}) differs from the one appearing in Eq.~(\ref{2.8}),  
where we approximated $a_i$ and $a_{i'}^{\dagger}$ as commuting. 
The non-vanishing commutation relations in Eq.~(\ref{a5}) 
explicitly yield a non-vanishing two-time commutator 
for ${{\cal O}}^{\Lambda}_i$. In particular, a straightforward calculation gives 
($i=1,2$): 
\beq
\label{a7}
C_{{\cal O}_i^\Lambda {\cal O}_i^\Lambda}(t,t')\equiv
\left[{{\cal O}}^{\Lambda}_i(t),{{\cal O}}^{\Lambda}_i(t')\right]=
i\frac{\Lambda^2}{\omega_o}
\left[\frac{\Lambda\tau\cos(\Lambda \tau)-\sin(\Lambda\tau)}{\pi\left(\Lambda\tau\right)^2}\right]\,,
\quad\quad \tau=t-t'\,.
\eeq
Therefore ${{\cal O}}^{\Lambda}_i(t)$ cannot be the final output and 
there must be some unavoidable additional quantum noise due to the fact 
that ${{\cal O}}^{\Lambda}_i(t)$ has a non-vanishing two-time commutator. 
In LIGO-I/TAMA/Virgo this additional noise is introduced in the output during the final 
process of photodetection. A more detailed study would involve a very technical 
analysis of the photodetection's dynamics, but fortunately, as we shall 
see in the following, a simple estimation of the order of magnitude of this 
additional quantum noise suggests that it is very small and we can realistically 
neglect it. 

We find it convenient to estimate the additional quantum noise by 
calculating the noise induced by the photodetector 
approximated as a linear measurement device coupled 
to the quadrature fields. \footnote{~Here we are assuming that as a consequence 
of the homodyne detection, the EM field impinging on the photodetector 
is composed of carrier light plus quantum fluctuations, 
and thus the light intensity measured by the photodetector 
is linear in the annihilation and creation operators.} 
Having fixed the cutoff frequency $\Lambda$ and working in the Fourier domain, 
we can write the final output as:
\beq
\label{a8}
{{\cal O}}^{out}_i(\Omega)={{\cal O}}^{\Lambda}_i(\Omega)+ 
{Z}^{{\rm PD}}_i(\Omega)+R_{{\cal O}_i^{\Lambda}{\cal O}_i^{\Lambda}}(\Omega)\,{F}^{{\rm PD}}_i(\Omega)\,,
\eeq
where 
\beq
\label{a9}
R_{{\cal O}_i^{\Lambda}{\cal O}_i^{\Lambda}}
(\Omega)\equiv\frac{i}{\hbar}\int_{0}^{+\infty}d\tau\,e^{i\Omega\tau}\,
C_{{\cal O}_i^\Lambda {\cal O}_i^\Lambda}(t,t-\tau) =
\frac{1}{2 \pi\hbar\omega_o}\left(2\Lambda+i\pi\Omega+\Omega\ln\frac{\Lambda-\Omega}{\Lambda+\Omega}\right)\,.
\eeq
The last two terms in Eq.~(\ref{a8}) are the shot noise and the 
back-action noise of the photodetector (PD) 
and describe the efficiency and the strength of perturbation of the PD on the 
quadrature field, respectively. Let us assume that there is no 
correlation between ${Z}^{{\rm PD}}_i$ and ${F}^{{\rm PD}}_i$.
Hence, ${Z}^{{\rm PD}}_i$ and ${F}^{{\rm PD}}_i$ satisfy 
the uncorrelated version (\ref{3.16}) of the uncertainty relation, that is 
\beq
\label{a10}
S_{{Z}_i^{{\rm PD}}{Z}_i^{{\rm PD}}}S_{{F}_i^{{\rm PD}} 
{F}_i^{{\rm PD}}} \ge \hbar^2\,.
\eeq
We are interested in evaluating the overall quantum noise. 
We first write the output in the form Signal + Noise as
\beq
\label{a11}
{{\cal O}}^{\rm out}_i(\Omega)={\cal{P}}_i h(\Omega) + 
\left[{\cal Q}^{\Lambda}_i(\Omega)
 +{Z}^{{\rm PD}}_i(\Omega)+R_{{\cal O}_i^{\Lambda}{\cal O}^{\Lambda}_i}
(\Omega){F}^{{\rm PD}}_i(\Omega)\right]\,,
\eeq
where ${\cal{P}}_i h$ is the part of ${{\cal O}}^{\Lambda}_i(\Omega)$ 
that contains the signal, while 
${\cal Q}^{\Lambda}_i(\Omega)$ contains the quantum fluctuations. 
Using Eq.~(\ref{a11}), the overall noise spectral density is ($i =1,2$):
\beq
\label{a12}
S_{i}(\Omega)=\frac{1}{\left|{\cal P}_i\right|^2}\left\{
S_{{\cal Q}_i^{\Lambda}{\cal Q}_i^{\Lambda}}(\Omega)+
S_{{Z}_i^{{\rm PD}}{Z}_i^{{\rm PD}}}(\Omega)+
\left|R_{{\cal O}_i^{\Lambda}{\cal O}^{\Lambda}_i}
(\Omega)\right|^2 S_{{F}_i^{{\rm PD}}{F}_i^{{\rm PD}}}(\Omega)\right\}\,.
\eeq
The first term in Eq.~(\ref{a12}) describes the quantum 
noise of an interferometer when the non-vanishing 
commutators of the quadrature fields have been ignored 
and ideal photodetection is applied. 
The second term in Eq.~(\ref{a12}) describes the additional shot noise 
introduced by the photodetection process. 
Finally, the third term comes from the back-action force acting on the 
measured quadrature ($i = 1$ or $2$) because it does 
not commute with itself at different times. 
Let us notice that, given Eq.~(\ref{a10}), the second and third noise contributions 
appearing on the RHS of Eq.~(\ref{a12})
are complementary. Indeed, the larger the shot noise,
the weaker the minimum force the photodetector must apply 
to the quadrature fields and the smaller 
the back-action noise. More specifically, there is a lowest achievable value 
for the PD part in Eq.~(\ref{a12}) given by:
\bea
\frac{1}{|{\cal P}_i|^2}\left[
S_{{Z}_i^{{\rm PD}} {Z}_i^{{\rm PD}}}(\Omega)+
\left|R_{{\cal O}^{\Lambda}_i{\cal O}^{\Lambda}_i}(\Omega)\right|^2 
S_{{F}_i^{{\rm PD}}{F}_i^{{\rm PD}}}(\Omega)\right] &\ge& 
\frac{2 |R_{{\cal O}^{\Lambda}_i{\cal O}^{\Lambda}_i}(\Omega)| \hbar}{|{\cal P }_i|^2}\,, 
\nonumber \\
&=&\frac{2}{|{\cal P }_i|^2}
\left|\frac{\Lambda}{\pi\omega_o} 
\left(1+\frac{\Omega}{2\Lambda}\ln\frac{\Lambda-\Omega}{\Lambda+\Omega}\right)
+i\frac{\Omega}{\omega_o}\right|\,.
\label{a13}
\eea
Using Eq.~(\ref{3.8}) we derive ${1}/{|{\cal S}_i|^2}={h_{\rm SQL}^2}/{2 \cal K}$ and 
$S_{{\cal Q}_i^{\Lambda}{\cal Q}_i^{\Lambda}} = ({\cal K}^2 + 1) > 1$. Recalling
that $10\,{\rm Hz}\le \Omega/2 \pi \le10^3 \,{\rm Hz}$ and $\omega_o 
\! \sim \! 10^{15}\,{\rm sec}^{-1}$, 
fixing $\Lambda$ to a value larger than the typical 
$\Omega$, e.g., $\Lambda/2 \pi \! \sim \! 10\,{\rm MHz}$, and adjusting the PD such that 
$S_{{Z}_i^{{\rm PD}} {Z}_i^{{\rm PD}}}$ and $S_{{F}_i^{{\rm PD}} {F}_i^{{\rm PD}}}$
satisfy the minimal uncertainty relation [the equality sign in Eq.~(\ref{a10})], 
we find that the minimal achievable PD noise
is $\sim \! 10^{-7}$ times the conventional shot noise. Therefore, 
we can totally ignore the quantum noise introduced by the fact that 
the quadrature fields do not commute with themselves at different 
times in Eq.~(\ref{a5}). Note the importance of the cutoff $\Lambda$. 
If we had taken $\Lambda \! \sim \! \omega_o$, the limit on the PD 
noise would have been of the same order of magnitude as 
the shot noise for a conventional interferometer and  
it would not have been realistic to neglect the quantum 
noise  introduced by the non-vanishing commutation relations of the quadrature fields.

So far we evaluated the minimum quantum noise that the photodetector, coupled 
linearly to the quadrature field, can introduce. Let us now try  
to give a realistic value of it. To estimate $S_{{Z}_i^{{\rm PD}}{Z}_i^{{\rm PD}}}$, 
we can just recall that in the case of a lossy photodetector we have (see the discussion 
of lossy interferometers in Sec.~\ref{sec5})
\beq
\label{a14}
{Z}^{{\rm PD}}_i \! \sim \! \sqrt{\lambda_{{\rm PD}}}\,{p}_i\,,
\eeq
where $p_i$ with $i = 1,2$ are quadrature operators in the 
vacuum state. We expect $\lambda_{{\rm PD}} \! \sim \! 0.1-0.2$, 
hence $S_{{Z}_i^{{\rm PD}}{Z}_i^{{\rm PD}}} > 10^{-2} \times S_{\rm
shot\,noise}^{\rm conv}$, 
which is five orders of magnitude larger than the lowest achievable limit 
discussed above with  $\Lambda/2 \pi \! \sim \! 10\,{\rm MHz}$. 
Therefore, if one can justify fixing the cutoff $\Lambda /2\pi$ at 
$10\,{\rm MHz}$, and if the uncertainty relation 
(\ref{a10}) is satisfied with the equality sign, then 
one can conclude that the inefficiency will dominate over the 
minimum possible back-action noise by five orders of 
magnitude. Hence, we are justified in disregarding the 
non-vanishing two-time commutators of the quadrature fields in 
Eq.~(\ref{a5}).

\end{document}